\documentclass{article}

\PassOptionsToPackage{table}{xcolor}

\usepackage[final,eandd]{neurips_2026}

\usepackage[utf8]{inputenc}
\usepackage[T1]{fontenc}
\usepackage[hidelinks]{hyperref}
\usepackage{url}
\usepackage{booktabs}
\usepackage{amsfonts}
\usepackage{nicefrac}
\usepackage{microtype}
\usepackage[table]{xcolor}

\usepackage{soul}
\usepackage{graphicx}
\usepackage{amsmath}
\usepackage{amsthm}
\usepackage{booktabs}
\usepackage{algorithm}
\usepackage{algorithmic}
\usepackage{mdframed}
\usepackage[small]{caption}

\newcommand{\adshotrepo}{\href{https://github.com/ostadabbas/AdShot}{\texttt{github.com/ostadabbas/AdShot}}}
\newcommand{\adshotdata}{\href{https://huggingface.co/datasets/omrastogi/adshot_benchmark}{\texttt{huggingface.co/datasets/omrastogi/adshot\_benchmark}}}

\definecolor{aboveavg}{RGB}{217,234,211}
\newcommand{\hi}[1]{\cellcolor{aboveavg}#1}

\usepackage{amssymb}
\usepackage{multicol}
\usepackage{multirow}
\usepackage{threeparttable}
\usepackage{bigdelim}
\usepackage{array}

\renewcommand{\arraystretch}{0.95}

\newcommand{\figyoutube}{
\begin{figure}[t]
    \centering
    \includegraphics[
        width=1\linewidth,
        trim=1cm 8.8cm 10cm 0.5cm, clip
    ]{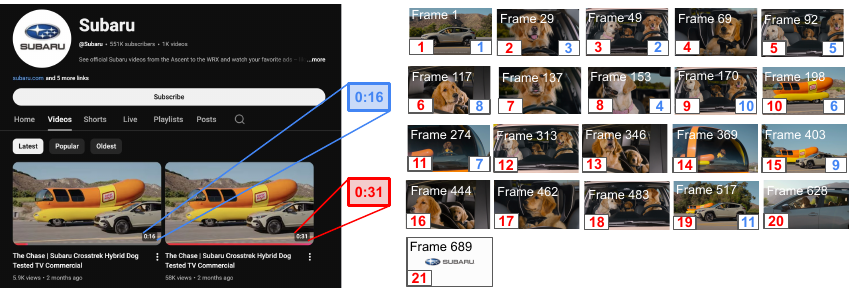}
    \caption{Example of the shot selection task: From a 31-sec Subaru ad from YouTube (shots indexed in red on the right panel; we show the first frame of each shot for better representation), professional editors select and order shots (blue indices) to create a 16-sec version promoting the same product (CrossTrek).
    }
    \vspace{-0.1in}
    \label{fig:f0-youtube}
    \vspace{-0.1in}
\end{figure}
}

\newcommand{\figtwostages}{
\begin{figure*}[htp]
  \centering
  \includegraphics[width=1\linewidth,
        trim=0.28cm 2cm 0.25cm 5.6cm,
        clip]{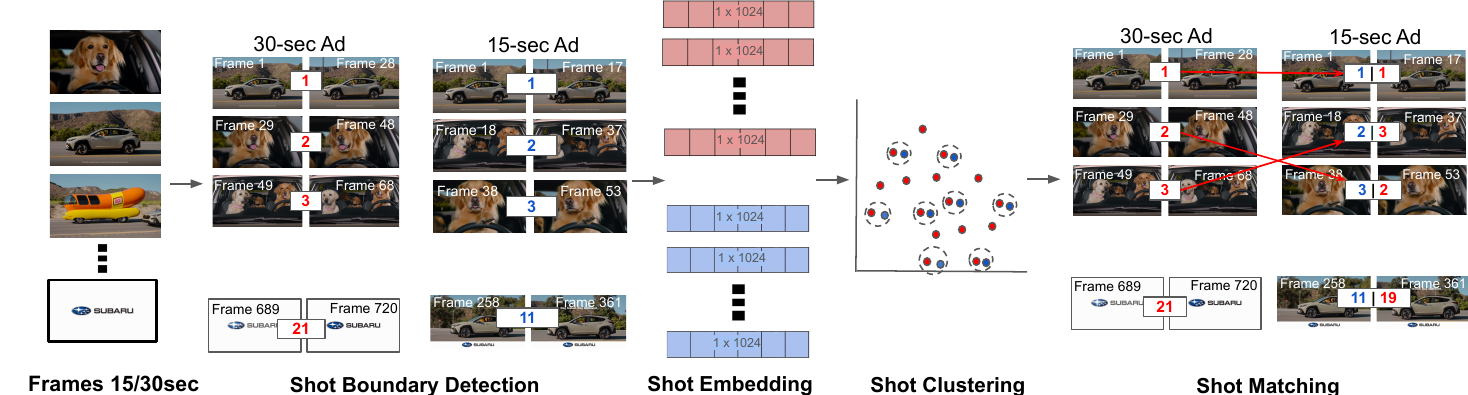}
  \caption{Methodological overview of long-short ad pair construction. Given videos from a brand's YouTube channel, we first apply AutoShot~\citep{zhuautoshot} for shot boundary detection, then extract shot embeddings using Swin3D~\citep{liu2022swin3d}. We cluster the embeddings using DBSCAN to match each shot in the 15-sec ad to its corresponding shot in the 30-sec ad. Throughout the figure, blue (red) elements (e.g., digits and dots) represent the 15-sec (30-sec) ad.   \label{fig:dataset}}
   \vspace{-.2in}
\end{figure*}
}

\newcommand{\figevaluation}{
\begin{figure}[H]
    \centering
\includegraphics[width=0.8\linewidth,
        trim=0cm 4.5cm 8cm 0cm,
        clip]{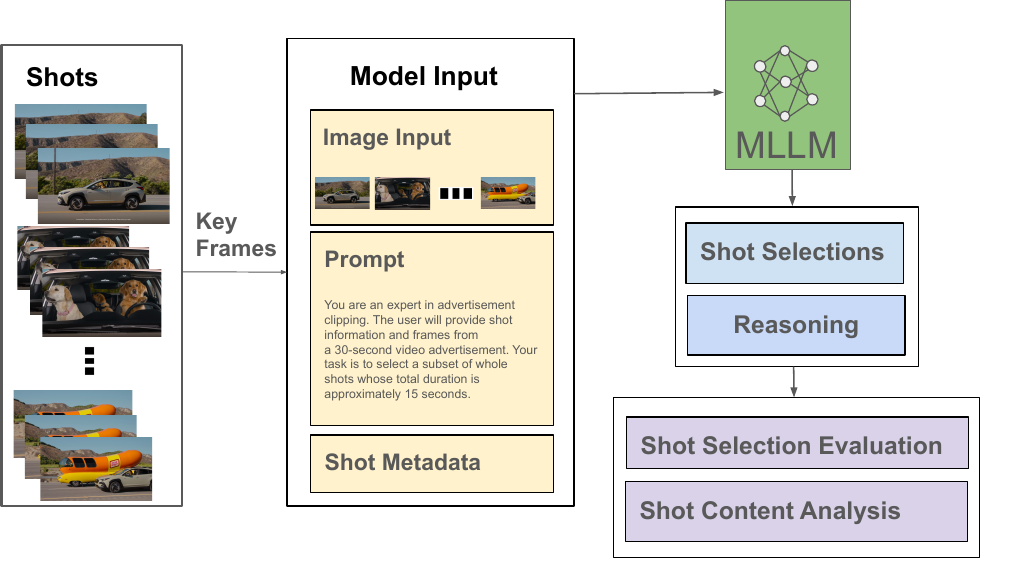}
  \caption{MLLM evaluation pipeline for ad shot selection. Keyframes from shots of a 30-sec ad, along with metadata (e.g., durations) and task instructions specifying the clipping objective, are fed to the model. The model outputs selected shots with reasoning, which are evaluated for accuracy and content preservation.
  }
    \label{fig:evaluation}
    \vspace{-.1in}
\end{figure}
}

\newcommand{\figcontent}{
\begin{figure}[H]
    \centering
\includegraphics[width=0.9\linewidth]{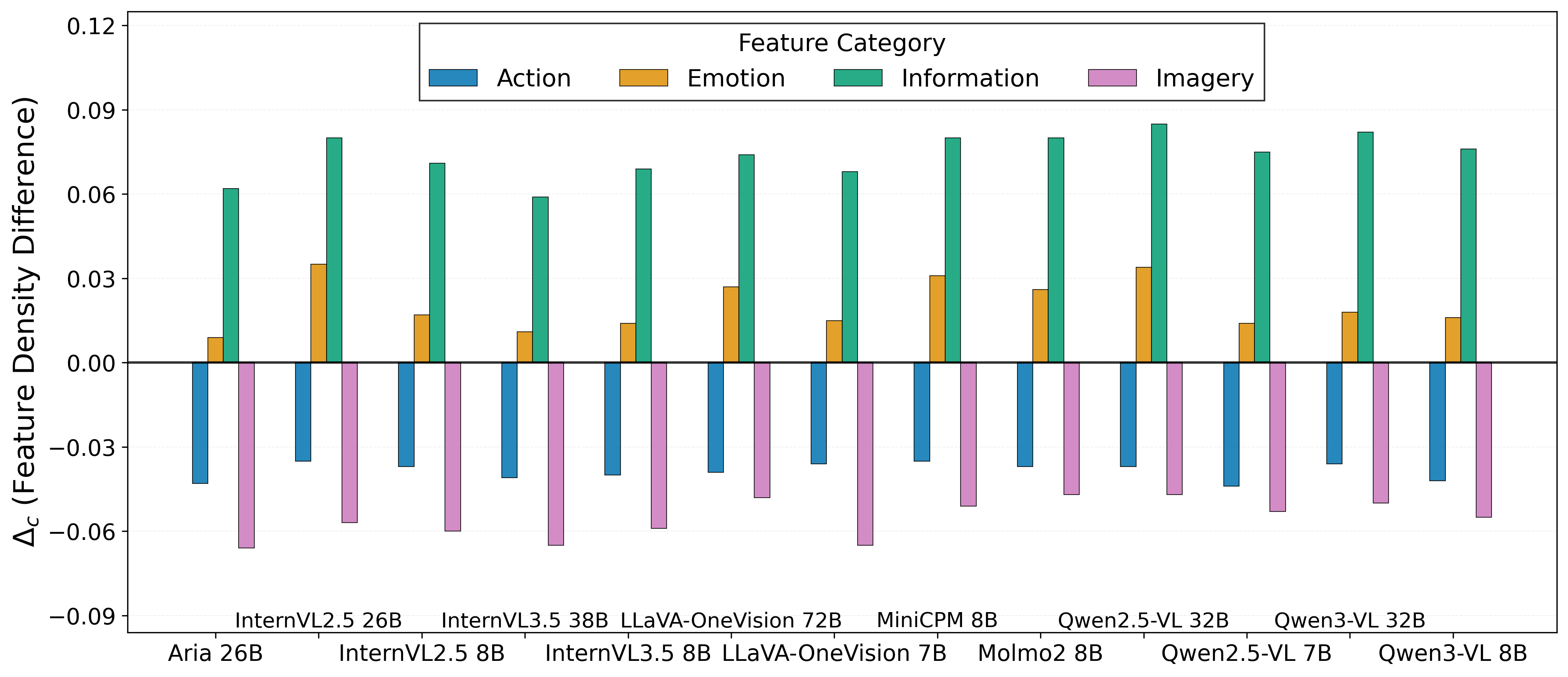}
  \caption{Content Feature Divergence. MLLMs over-select emotion- and information-focused shots while under-selecting action- and imagery-driven content, indicating systematic biases relative to professional edits. \label{fig:contentanalysis}}
    \vspace{-.1in}
\end{figure}
}

\newcommand{\figcomparison}{
\begin{figure}
    \centering
\includegraphics[width=0.9\linewidth]{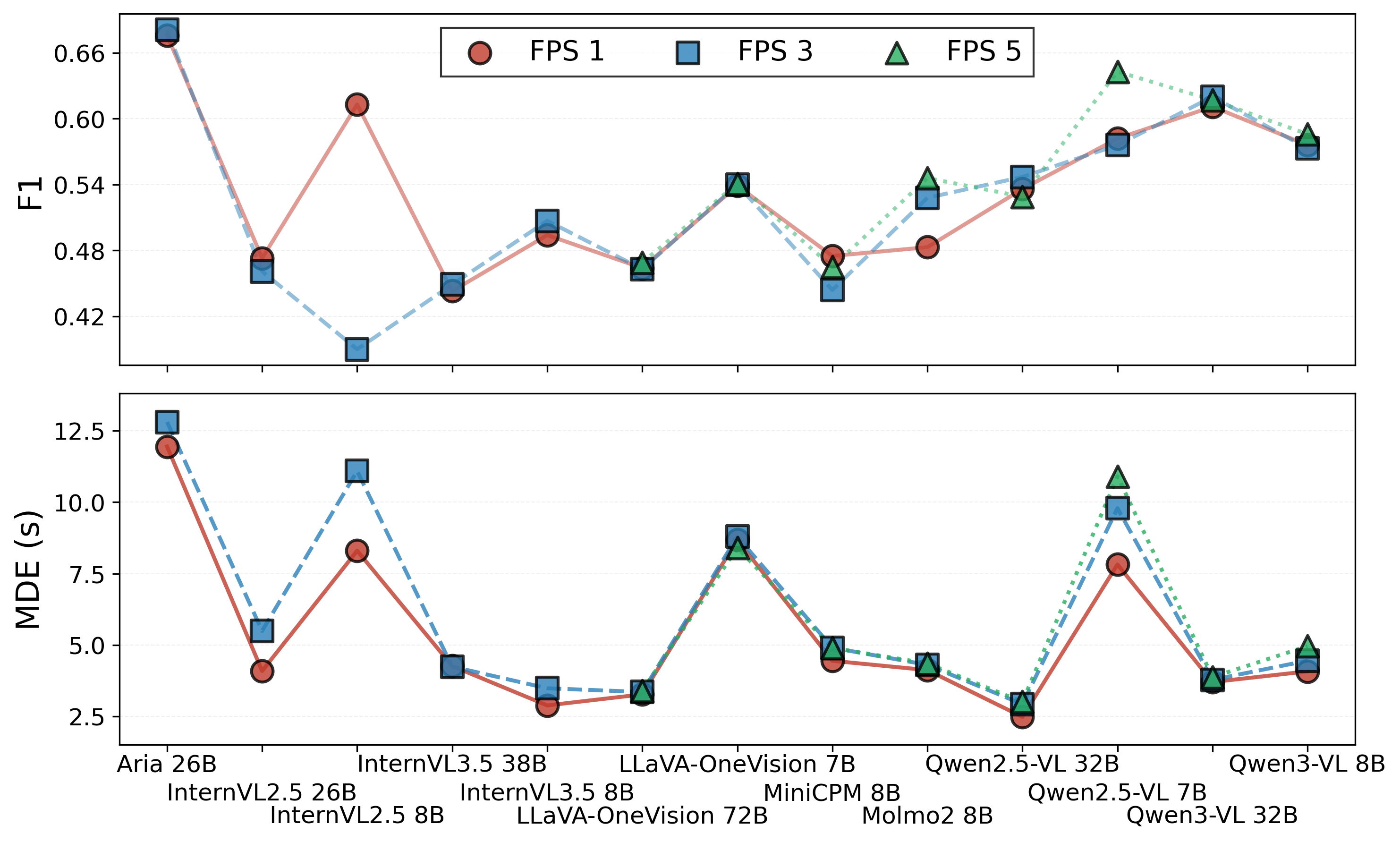}
  \caption{Ablation on temporal sampling. Using 3 or 5 frames per shot instead of 1 FPS does not consistently improve MLLM performance, indicating limited benefit from denser temporal input. \label{fig:allcomparision}}
    \vspace{-.1in}
\end{figure}
}
\usepackage{amssymb}
\usepackage{multicol}
\usepackage{multirow}
\usepackage{threeparttable}
\usepackage{bigdelim}
\usepackage{array}
\usepackage{makecell}
\usepackage[table]{xcolor}
\usepackage{booktabs}
\usepackage{caption}

\definecolor{higlight}{RGB}{220,235,255} 

\renewcommand{\arraystretch}{0.95}

\newcommand{\tabdataset}{
\begin{table}[t]
\centering
\small
\resizebox{\columnwidth}{!}{
\begin{tabular}{lccccl}
\toprule
\textbf{Benchmark} & \textbf{Domain} & \textbf{Task} & \textbf{Videos} & \textbf{Length} & \textbf{Key Distinction} \\
\midrule
VideoAds~\cite{zhang2025videoads}                  & Ads     & Open-ended QA      & 3,153  & 15--60s    & QA on ad content \\
AdsQA~\cite{long2025adsqa}                         & Ads     & Open-ended QA      & 600    & 5--90s     & Reasons about implicit ad meanings \\

AdSum~\citet{xie2026adsum}  & Ads & Shot Selection & 204 & 15s, 30s & Supervised audio-visual model; no MLLM evaluation \\

Video-MME~\cite{fu2025video}                       & General & Multiple-choice QA & 900    & 11s--1h    & Broad multimodal understanding \\
MVBench~\cite{li2023mvbench}                       & General & Open-ended QA      & 4,000  & 10s--10min & Temporal reasoning across events \\
MLVU~\cite{zhou2025mlvubenchmarkingmultitasklong}  & General & Multiple-choice QA & 1,500+ & Hours      & Long-form video reasoning \\

\midrule
\textbf{AdShot (Ours)} & \textbf{Ads} & \textbf{Shot Selection} & \textbf{1,646} & \textbf{15s, 30s} & \textbf{Shot selection for ad clipping} \\
\bottomrule
\end{tabular}
}
\caption{Comparison between AdShot and existing video understanding benchmarks.}
\label{tab:benchmark_comparison}
\vspace{-0.2cm}
\end{table}
}

\newcommand{\tabdatasetstats}{
\begin{figure*}[t]
\centering
\begin{minipage}[t]{0.37\linewidth}
\vspace{0pt}
\centering
\includegraphics[width=\linewidth]{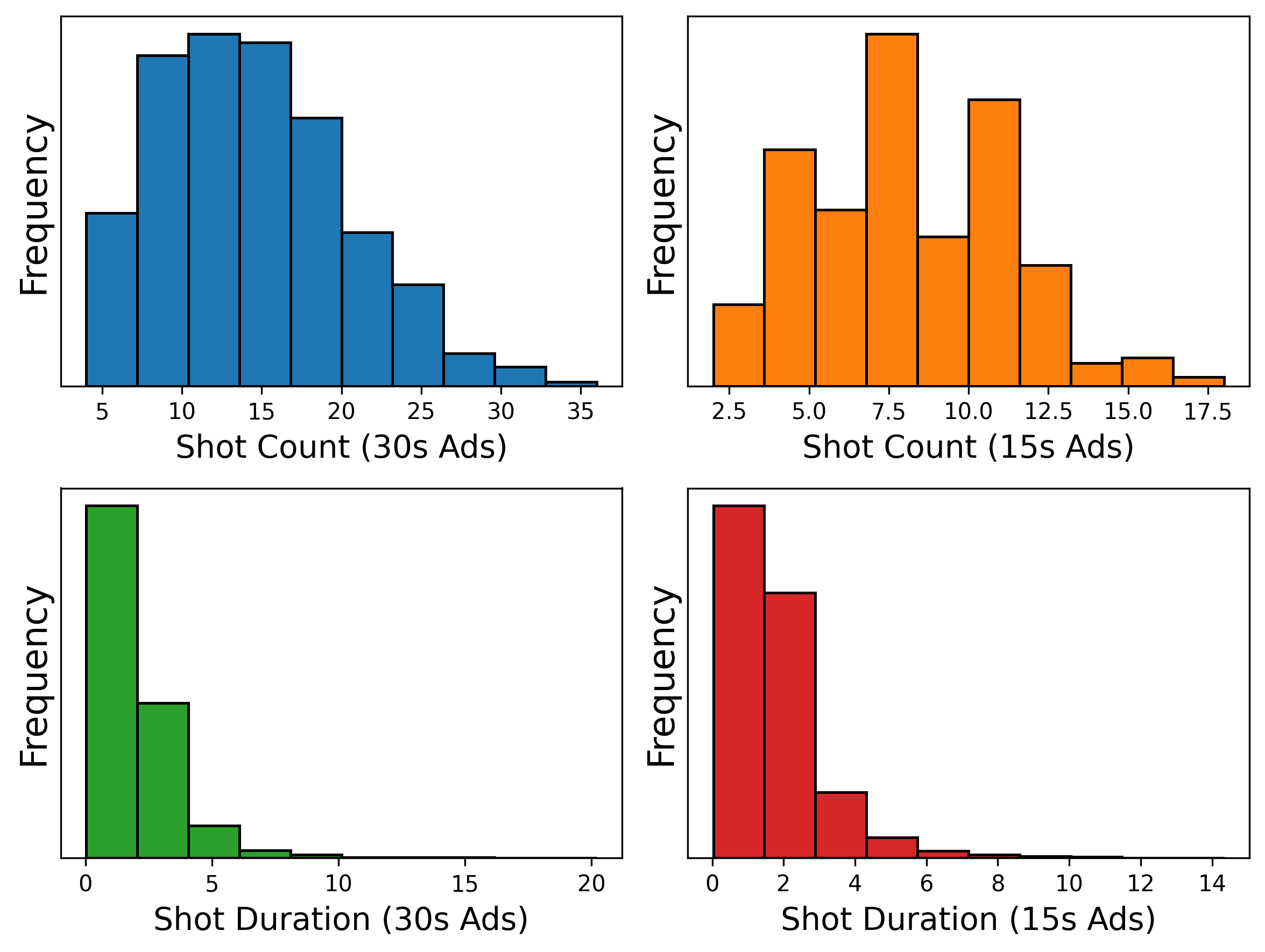}
\captionof{figure}{Histograms of shot counts and durations. On average, 30-sec (15-sec) ads contain 14 (8) shots. The average shot duration is 2.05 seconds for 30-sec ads and 1.83 seconds for 15-sec ads.}
\label{fig:shotstats}
\end{minipage}
\hfill
\begin{minipage}[t]{0.60\linewidth}
\vspace{0pt}
\centering
\vspace{0.2em}
\small
\resizebox{\linewidth}{!}{
\begin{tabular}{lcccccc}
\toprule
\multirow{2}{*}{\textbf{Industry}} & \multirow{2}{*}{\textbf{\# Brands}} & \multirow{2}{*}{\textbf{\# Ad Pairs}} & \textbf{Shots} & \textbf{Avg. Shots/Ad} & \textbf{Avg. Dur./Ad} \\
 & & & (30s / 15s) & (30s / 15s) & (30s / 15s) \\
\midrule
Apparel            & 3  & 4   & 80 / 46     & 20.00 / 11.50 & 1.50 / 1.30 \\
Automotive         & 15 & 44  & 767 / 435   & 17.43 / 9.89  & 1.72 / 1.51 \\
Consumer Products  & 9  & 23  & 345 / 188   & 15.00 / 8.17  & 2.00 / 1.80 \\
Cosmetic           & 5  & 15  & 215 / 109   & 14.33 / 7.27  & 2.10 / 2.05 \\
Education          & 4  & 27  & 444 / 236   & 16.44 / 8.74  & 1.82 / 1.71 \\
Electronics        & 2  & 2   & 35 / 21     & 17.50 / 10.50 & 1.70 / 1.41 \\
Entertainment      & 6  & 17  & 310 / 115   & 18.24 / 6.76  & 1.63 / 2.25 \\
Food \& Beverages  & 18 & 40  & 642 / 383   & 16.05 / 9.57  & 1.87 / 1.57 \\
Government         & 9  & 108 & 1400 / 786  & 12.96 / 7.28  & 2.31 / 2.07 \\
Health             & 19 & 45  & 611 / 346   & 13.58 / 7.69  & 2.20 / 1.95 \\
Home and Garden    & 10 & 17  & 281 / 158   & 16.53 / 9.29  & 1.82 / 1.61 \\
Legal              & 25 & 125 & 1507 / 849  & 12.06 / 6.79  & 2.48 / 2.20 \\
Restaurants        & 6  & 22  & 292 / 160   & 13.27 / 7.27  & 2.25 / 2.05 \\
Retail             & 32 & 159 & 2462 / 1497 & 15.48 / 9.42  & 1.93 / 1.59 \\
Services           & 19 & 115 & 1698 / 903  & 14.77 / 7.85  & 2.03 / 1.92 \\
Telecommunications & 4  & 37  & 521 / 283   & 14.08 / 7.65  & 2.12 / 1.95 \\
Travel             & 8  & 23  & 388 / 214   & 16.87 / 9.30  & 1.78 / 1.61 \\
\bottomrule
\end{tabular}
}
\captionof{table}{AdShot dataset statistics by industry. Shots and durations are reported for 30-sec and 15-sec ad versions.}
\label{tab:dataset_stats}
\vspace{-0.5em}
\end{minipage}
\end{figure*}
}

\newcommand{\tabperformance}{
\begin{table}[t]
\centering
\vspace{0.2em}
\footnotesize
\renewcommand{\arraystretch}{0.85}
\resizebox{0.85\columnwidth}{!}{
\begin{tabular}{lcccccc}
\toprule
\textbf{Model} & \textbf{Precision $\uparrow$} & \textbf{Recall $\uparrow$} & \textbf{F1 $\uparrow$} & \textbf{MDE $\downarrow$} & \textbf{MDE$_{\text{pos}}$ $\downarrow$} & \textbf{MDE$_{\text{neg}}$ $\downarrow$} \\
\midrule
Random baseline      & 0.569 & 0.541 & 0.503 & 8.18 & 9.01 & 7.16 \\
\midrule
\multicolumn{7}{l}{\textit{Video-only inputs}} \\
\midrule
Qwen2.5-VL 7B       & 0.613          & 0.649          & 0.582          & 7.82           & 5.92          & 1.90          \\
LLaVA-OneVision 7B  & 0.578          & 0.601          & 0.539          & 8.68           & 6.41          & 2.27          \\
Qwen3-VL 8B         & 0.618          & 0.580          & 0.576          & 4.08           & 2.99          & 1.10          \\
InternVL2.5 8B      & 0.594          & 0.698          & 0.613          & 8.30           & 7.98          & 0.32          \\
InternVL3.5 8B      & 0.614          & 0.432          & 0.494          & 2.89           & 0.47          & 2.42          \\
MiniCPM 8B          & 0.599          & 0.414          & 0.475          & 4.45           & 1.20          & 3.25          \\
Molmo2 8B           & 0.602          & 0.450          & 0.483          & 4.13           & 1.69          & 2.43          \\
\midrule
InternVL2.5 26B     & 0.614          & 0.411          & 0.473          & 4.10           & 1.94          & 2.15          \\
Aria 25.3B            & 0.576          & \textbf{0.881} & \textbf{0.676} & 11.95          & 11.76         & 0.18          \\
Qwen3-VL 32B        & \textbf{0.636} & 0.618          & 0.611          & 3.71           & 3.31          & 0.39          \\
Qwen2.5-VL 32B      & 0.626          & 0.490          & 0.536          & \textbf{2.49}  & 1.59          & 0.90          \\
InternVL3.5 38B     & 0.597          & 0.366          & 0.443          & 4.27           & \textbf{0.09} & 4.18          \\
LLaVA-OneVision 72B & 0.598          & 0.400          & 0.464          & 3.28           & 0.76          & 2.52          \\
\midrule
\multicolumn{7}{l}{\textit{Audio-visual inputs}} \\
\midrule
Qwen2.5-Omni 7B     & 0.615          & 0.838          & 0.668          & 11.18          & 12.68         & 3.16          \\
MiniCPM-o 2.6 8B    & 0.581          & 0.841          & 0.669          & 10.06          & 10.50         & \textbf{1.72} \\
\bottomrule
\end{tabular}
}
\caption{Performance on shot selection (1 FPS). Random baseline samples a random subset size $N$, then picks $N$ shots uniformly at random (Appendix~\ref{app:random_baseline}). Bold marks the best per column across all evaluated models. $\uparrow$/$\downarrow$ indicate higher/lower is better.}
\label{tab:main_results}
\vspace{-0.8em}
\end{table}
}

\newcommand{\tabfpsablation}{
\begin{table*}[tp]
\centering
\small
\begin{tabular}{lcccccc}
\toprule
\textbf{Model} &
\multicolumn{2}{c}{\textbf{fps=1}} &
\multicolumn{2}{c}{\textbf{fps=3}} &
\multicolumn{2}{c}{\textbf{fps=5}} \\
\cmidrule(lr){2-3}\cmidrule(lr){4-5}\cmidrule(lr){6-7}
& \textbf{F1 $\uparrow$} & \textbf{MDE $\downarrow$} & \textbf{F1 $\uparrow$} & \textbf{MDE $\downarrow$} & \textbf{F1 $\uparrow$} & \textbf{MDE $\downarrow$} \\
\midrule
\multicolumn{7}{l}{\textit{Video-only inputs}} \\
\midrule
Qwen2.5-VL 7B          & 0.582 & \cellcolor{higlight}7.82  & 0.576 & 9.79  & \cellcolor{higlight}0.643 & 10.91 \\
LLaVA-OneVision 7B     & 0.539 & 8.68  & 0.540 & 8.80  & \cellcolor{higlight}0.541 & \cellcolor{higlight}8.41  \\
Qwen3-VL 8B            & 0.576 & \cellcolor{higlight}4.08  & 0.573 & 4.46  & \cellcolor{higlight}0.586 & 4.96  \\
InternVL2.5 8B         & \cellcolor{higlight}0.613 & \cellcolor{higlight}8.30  & 0.390 & 11.09  & ---   & ---    \\
InternVL3.5 8B         & 0.494 & \cellcolor{higlight}\textbf{2.89}  & \cellcolor{higlight}0.507 & \textbf{3.49}  & ---   & ---    \\
MiniCPM 8B             & \cellcolor{higlight}0.475 & \cellcolor{higlight}4.45  & 0.444 & 4.91  & 0.465 & 4.90  \\
Molmo2 8B              & 0.483 & \cellcolor{higlight}4.13  & 0.528 & 4.31  & \cellcolor{higlight}0.546 & \textbf{4.36}  \\
\midrule
InternVL2.5 26B        & \cellcolor{higlight}0.473 & \cellcolor{higlight}4.10  & 0.461 & 5.50  & ---   & ---    \\
Aria 25.3B               & 0.676 & \cellcolor{higlight}11.95 & \cellcolor{higlight}\textbf{0.681} & 12.81 & ---   & ---    \\
Qwen3-VL 32B           & 0.611 & \cellcolor{higlight}3.71  & \cellcolor{higlight}0.620 & 3.79  & \textbf{0.617} & 3.88  \\
Qwen2.5-VL 32B         & 0.536 & \cellcolor{higlight}\textbf{2.49}  & \cellcolor{higlight}0.547 & \textbf{2.94}  & 0.529 & \textbf{3.01}  \\
InternVL3.5 38B        & 0.443 & 4.27  & \cellcolor{higlight}0.449 & \cellcolor{higlight}4.25  & ---   & ---    \\
LLaVA-OneVision 72B    & 0.464 & \cellcolor{higlight}3.28  & 0.463 & 3.36  & \cellcolor{higlight}0.469 & 3.40  \\
\midrule
\multicolumn{7}{l}{\textit{Audio + Video inputs}} \\
\midrule
Qwen2.5-Omni 7B        & \cellcolor{higlight}0.668 & 11.18 & 0.623 & \cellcolor{higlight}10.26 & 0.624 & 10.40 \\
MiniCPM-o 2.6 8B       & \cellcolor{higlight}\textbf{0.669} & 10.06 & 0.590 & \cellcolor{higlight}8.17  & 0.617 & 8.56  \\
\bottomrule
\end{tabular}
\caption{Ablation on FPS. Shot selection performance under different temporal sampling densities. $\uparrow$/$\downarrow$ indicate higher/lower is better. Shaded cells highlight, for each model, the best performance across the three FPS values; bold marks the best across all models. Some models (e.g., InternVL2.5) are excluded from 5 FPS due to token-length constraints.}
\label{tab:fps_performance}
\end{table*}
}

\newcommand{\tabshotbins}{
\begin{table*}[tp]
\centering

\vspace{-0.5em}
\footnotesize
\renewcommand{\arraystretch}{0.95}
\setlength{\tabcolsep}{10pt}
\begin{tabular*}{\textwidth}{l@{\extracolsep{\fill}}cccccc}
\toprule
\textbf{Model} &
\multicolumn{3}{c}{\textbf{F1 $\uparrow$}} &
\multicolumn{3}{c}{\textbf{MDE (s) $\downarrow$}} \\
\cmidrule(lr){2-4}\cmidrule(lr){5-7}
& \textbf{SC$<$6} & \textbf{6$\leq$SC$\leq$10} & \textbf{SC$>$10} & \textbf{SC$<$6} & \textbf{6$\leq$SC$\leq$10} & \textbf{SC$>$10} \\
\midrule
\multicolumn{7}{l}{\textit{Video-only inputs}} \\
\midrule
Qwen2.5-VL 7B       & 0.609                       & \textbf{0.706}              & \cellcolor{higlight}0.719          & \cellcolor{higlight}9.95           & 12.44                      & 14.05          \\
LLaVA-OneVision 7B  & \cellcolor{higlight}0.534   & 0.467                       & 0.381                              & 4.46                               & \cellcolor{higlight}3.40   & 5.21           \\
Qwen3-VL 8B         & 0.595                       & 0.614                       & \cellcolor{higlight}0.640          & 7.39                               & 8.11                       & 10.40          \\
InternVL2.5 8B      & \cellcolor{higlight}0.480   & 0.450                       & 0.360                              & \cellcolor{higlight}3.32           & 4.06                       & 6.45           \\
InternVL3.5 8B      & 0.503                       & \cellcolor{higlight}0.515   & 0.427                              & \textbf{2.54}                      & \cellcolor{higlight}2.52   & 4.44           \\
MiniCPM 8B          & \cellcolor{higlight}0.494   & 0.454                       & 0.436                              & \cellcolor{higlight}2.96           & 3.22                       & 3.99           \\
Molmo2 8B           & 0.532                       & 0.518                       & \cellcolor{higlight}0.604          & \cellcolor{higlight}7.92           & 8.14                       & 11.41          \\
\midrule
InternVL2.5 26B     & \cellcolor{higlight}0.517   & 0.473                       & 0.401                              & 4.27                               & \cellcolor{higlight}4.06   & 5.83           \\
Aria 25.3B            & \cellcolor{higlight}0.498   & 0.468                       & 0.495                              & \cellcolor{higlight}3.14           & 4.18                       & 5.73           \\
Qwen3-VL 32B        & 0.537                       & \cellcolor{higlight}0.546   & 0.508                              & 2.71                               & \cellcolor{higlight}\textbf{1.98} & \textbf{3.41} \\
Qwen2.5-VL 32B      & 0.543                       & 0.597                       & \cellcolor{higlight}0.612          & \cellcolor{higlight}5.19           & 8.61                       & 10.43          \\
InternVL3.5 38B     & \cellcolor{higlight}0.614   & 0.611                       & 0.606                              & 4.10                               & \cellcolor{higlight}3.46   & 3.66           \\
LLaVA-OneVision 72B & 0.546                       & 0.598                       & \cellcolor{higlight}0.569          & \cellcolor{higlight}3.55           & 3.91                       & 5.45           \\
\midrule
\multicolumn{7}{l}{\textit{Audio-visual inputs}} \\
\midrule
Qwen2.5-Omni 7B     & \cellcolor{higlight}\textbf{0.716} & 0.656                & 0.672                              & \cellcolor{higlight}8.75           & 11.42                      & 11.99          \\
MiniCPM-o 2.6 8B    & 0.580                       & 0.686                       & \cellcolor{higlight}\textbf{0.717} & \cellcolor{higlight}7.60           & 10.13                      & 12.77          \\
\bottomrule
\end{tabular*}
\caption{Shot selection performance by targeted shot count (SC) at fps=1. F1 and MDE for 15-sec ads with $<$6, 6--10, and $>$10 shots. Shaded cells highlight per-model best F1 and lowest MDE across bins; bold marks the best per column.}
\label{tab:shotbin_results}
\vspace{-0.8em}
\end{table*}
}

\title{AdShot: Benchmarking Multimodal Large Language Models for Video Advertisement Clipping}

\author{%
  Wen Xie$^{1}$ \quad
  Om Rastogi$^{1}$ \quad
  Sai Siddhartha Vivek Dhir Rangoju$^{1}$ \\
  Gijs Overgoor$^{2}$ \quad
  Yakov Bart$^{1}$ \quad
  Sarah Ostadabbas$^{1}$\thanks{Correspondence to \texttt{s.ostadabbas@northeastern.edu}.} \\[0.6em]
  $^{1}$Northeastern University, Boston, MA 02115, USA \\
  $^{2}$Southern Methodist University, Dallas, TX 75205, USA
}

\begin{document}

\maketitle

\begin{abstract}
Advertisers routinely create multiple duration variants of ads to meet marketing budgets and viewer preferences, a labor-intensive and costly process. While multimodal large language models (MLLMs) excel at general video understanding, their ability to perform specialized editorial tasks remains unexplored. We introduce \textbf{AdShot}, a benchmark for evaluating MLLMs on shot selection for ad clipping. It contains 823 professionally edited ad pairs (30-sec sources and 15-sec edits) across 194 brands and 17 industries, annotated with shot-level boundaries and content features (action, emotion, information, imagery). We evaluate 13 state-of-the-art video-only MLLMs and 2 audio-visual MLLMs (7B–72B parameters) on shot selection accuracy (Precision, Recall, F1), duration constraints (15s), and content preservation
, including an audio ablation%
. We find: (1) model architecture matters more than scale, with smaller models often matching or exceeding larger ones; performance is strongest for target ads with 6–10 shots; (2) models exhibit a mean absolute duration error of 5.39s, with most tending to overshoot the 15s target, though they can still substantially reduce human editing effort; (3) all models exhibit systematic content biases, favoring information-rich and emotion-focused shots while under-selecting action and imagery. We publicly release our benchmark to facilitate research at the intersection of multimodal AI and computational advertising.

Codebase: \adshotrepo

Dataset: \adshotdata

\end{abstract}

\section{Introduction}
The video advertising market is rapidly expanding, with global spending projected to reach \$215 billion in 2025~\citep{statista2025}. To adapt to user attention spans and platform requirements, advertisers produce multiple duration variants of the same ad, most commonly 30-sec and 15-sec versions~\citep{mediaradar2023} (see Figure~\ref{fig:f0-youtube} for an example). The creation process involves manual shot selection and rearrangement from longer source ads, a labor-intensive editorial process that remains a major bottleneck in modern advertising production pipelines~\citep{xie2026adsum}.

Recent advances in multimodal large language models (MLLMs) have shown strong performance in video understanding, including video question answering~\citep{long2025adsqa}, temporal grounding~\citep{han2025videoespresso}, and long-form video analysis~\citep{fu2025video}. These results suggest that MLLMs can recognize narrative structure, identify salient events, and reason over temporal sequences.

However, video ads pose unique challenges beyond general video benchmarks. First, ad videos exhibit high temporal and informational density: a typical 30-sec ad contains 10–20 shots, often lasting only 1–3 seconds each~\citep{xie2026adsum}. This requires fine-grained temporal reasoning ~\citep{zhang2025videoads}. Second, ad content is purpose-driven. Each shot serves a strategic role, such as conveying product information, eliciting emotion, establishing imagery, or motivating action. Third, shot selection is subject to strict duration constraints (e.g., producing a 15-sec version from a 30-sec source). As a result, effective ad shot selection cannot simply select the most salient moments, but preserves a balanced content composition aligned with advertising objectives.

\figyoutube

To address this gap, we introduce \textbf{AdShot}, the first large-scale benchmark designed to evaluate MLLMs on shot selection for ad clipping. Our benchmark assesses whether current models can replicate professional editorial decisions made in real-world advertising. AdShot consists of 823 professionally produced ad pairs, each comprising a 30-sec source ad and its corresponding 15-sec version released by the same brand. The dataset spans 194 brands across 17 industries. All videos are annotated with precise shot boundaries, and each shot is labeled with high-level content attributes (action, emotion, information, and imagery) reflecting the elements that advertisers prioritize when crafting effective ads~\citep{liaukonyte2015television}. These annotations enable nuanced analysis of selection behavior.

Our work contributes to research at the intersection of multimodal AI and computational advertising~\citep{hussain2017automatic,xie2026adsum} in four key ways:
\begin{itemize}
\item Formalize ad clipping as a distinct multimodal editorial task, highlighting its unique challenges in temporal density, duration constraints, and content composition.
\item Introduce \textbf{AdShot}, a benchmark of 823 professionally edited ad pairs spanning 194 brands and 17 industries, enabling direct comparison between MLLM predictions and real-world professional editorial decisions.
\item Systematically evaluate 13 video-only and 2 audio-visual MLLMs, revealing that (1) architecture matters more than scale, (2) audio cues provide complementary signal with audio-visual models nearly matching the best video-only model at a third of the parameters, and (3) 
9 of the 15 models surpass a heuristic random baseline while 6 fall below it, so this ability is model-dependent rather than universal.%

\item Identify systematic content biases in MLLM shot selection, revealing consistent preferences for information- and emotion-driven content over action and imagery, with implications for real-world advertising use.
\end{itemize}

\section{Related Work}

\subsection{Video Understanding Benchmarks}
Video understanding benchmarks evaluate a model's ability to interpret actions, events, and scenes across spatial and temporal dimensions. Widely used benchmarks include Video-MME for comprehensive multimodal evaluation~\citep{fu2025video}, MVBench for temporal understanding~\citep{li2023mvbench}, and MLVU for long-form video reasoning~\citep{zhou2025mlvubenchmarkingmultitasklong}. Recognizing the economic significance and unique characteristics of video advertising, such as short shot durations and rapid content transitions, recent work has introduced ad-specific benchmarks. For example, VideoAds~\citep{zhang2025videoads}, and AdsQA~\citep{long2025adsqa} focus on understanding and reasoning about advertisement content. However, these benchmarks primarily assess comprehension through question answering. 

Most related to our work, \citet{xie2026adsum} frame video ad clipping as a shot 
selection problem and propose a two-stream audio-visual fusion model trained on a 
paired ad dataset (AdSum204). While that work demonstrates the value of domain-specific training and highlights the importance of audio cues, it relies on a supervised architecture 
with binary shot annotations and does not characterize the content properties of 
selected shots. In contrast, AdShot evaluates zero-shot MLLMs on editorial 
decision-making, measuring alignment with professional human edits under strict 
duration constraints, as summarized in Table~\ref{tab:benchmark_comparison}.

\vspace{-0.5cm}
\subsection{Video Summarization and Editing with MLLMs}
Video summarization condenses long videos into shorter representative clips~\citep{apostolidis2021video}, with domain-specific variants for surveillance~\citep{Fang2016Abnormal}, sports~\citep{Chen2025FootballActionRecognition,khan2020Context,santra2025preciseeventspottingsports}, movie trailers~\citep{wang2024inversepartialoptimaltransport,Yao2025AutomaticTrailer,argaw2024automatedmovietrailergeneration}, and news~\citep{Krubinski2023MLASK}. Ad clips differ fundamentally: shots are short and information-dense, and effective selection must preserve a balanced composition of product demonstrations, emotional appeals, and calls to action under strict duration constraints.

Emerging work has explored MLLMs for ad video generation: \citet{wang2025long} create engaging clips from long videos, and \citet{qian2025vc} propose VC-LLM for generating ads from raw footage. These approaches evaluate generation quality rather than alignment with professional edits, the gap AdShot directly addresses through paired 30-to-15-second references for systematic evaluation of editorial judgment. 
\tabdataset

\figtwostages

\vspace{0cm}
\subsection{Computational Advertising}
Marketing research establishes that effective ad design strategically employs four critical content dimensions to achieve distinct objectives: \textit{action} drives calls-to-action (e.g., visit website, contact, purchase), \textit{emotion} builds brand affinity and memory, \textit{information} conveys factual product claims and attributes, and \textit{imagery} establishes visual identity and aesthetic appeal~\citep{liaukonyte2015television}. The composition of these elements varies by campaign goals, e.g., brand awareness campaigns emphasize emotion and imagery, while direct-response ads prioritize information and action. These considerations motivate our evaluation approach. We further assess whether MLLMs preserve the content composition strategies that characterize professional editorial decisions.

\section{The AdShot Benchmark}
\label{sec:benchmark}
\subsection{Ad Selection Task Formulation}
\label{sec:task}
Given a 30-sec ad video $V_{30}$ segmented into $n$ temporally ordered shots
$\{s_1, s_2, \ldots, s_n\}$, where each shot $s_i$ has duration $d_i$ seconds and
\begin{equation}
\sum_{i=1}^{n} d_i = 30.
\end{equation}
Let $S^* \subseteq \{s_1, s_2, \ldots, s_m\}$ denote the ground-truth subset of shots selected by professional editors to construct the corresponding 15-sec ad video $V_{15}$.

\paragraph{Task.}
The objective is to select a subset of shots $S \subseteq \{s_1, s_2, \ldots, s_l\}$ that satisfies the duration constraint:
\begin{equation}
14 \leq \sum_{s_i \in S} d_i \leq 16,
\end{equation}
where $\sum_{s_i \in S} d_i$ denotes the total duration (seconds) of the selected shots. 

\paragraph{Objective.}
The selection $S$ should maximize overlap with the ground-truth $S^*$ derived from the professionally edited video $V_{15}$, while meeting the duration constraint.

This formulation reflects unique challenges of ad clipping: selecting a small set of shots that strategically preserve certain content composition under strict duration constraints.

\paragraph{Scope.}
AdShot measures shot selection, the labor-intensive core of a cutdown, at whole-shot granularity: ad shots are already short (10--20 per 30-sec ad, \S1), so a keep/drop decision per shot captures the editorial choice. The steps that follow selection are out of scope and unmeasured: reordering, intra-shot trimming, transitions, grading, and audio mixing. Reordering is in any case rare (45 of the 50 pairs in the manual review of \S\ref{sec:construction} preserve the original order exactly) but we treat order, flow, and splice quality as explicitly unmeasured (\S\ref{sec:limitations}).%


\subsection{Dataset Construction}
\label{sec:construction}
We build AdShot as a benchmark for evaluating MLLMs on the ad clipping task. Figure~\ref{fig:dataset} illustrates the pipeline, which consists of four stages transforming raw videos into paired 30-sec and 15-sec ads. We describe each stage below.

\textbf{Video collection.} We start with a list of thousands of major TV advertisers across 17 industries (e.g., food and beverage, entertainment, automotive, retailing), provided by an advertising measurement company. The list spans brands such as Subaru, Tide, Capital One, and Visionworks. For each brand, we collect the complete catalog from its official YouTube channels, focusing on brands that produce multiple duration variants for the same campaign. This yields an initial pool of $\sim$700,000 videos from 836 brand channels, which we filter by title and duration to obtain $\sim$120,000 candidates.

\textbf{Shot boundary detection.} We employ AutoShot~\citep{zhuautoshot}, which was pretrained on over 11,000 shots with average duration (2.59s/shot) close to ours (2.12s/shot, see Table~\ref{tab:dataset_stats}). Following the literature \citep{xie2026adsum}, we set the detection threshold to 0.1 to capture fine-grained shot boundaries, applied consistently across 30-sec and 15-sec ads to enable reliable matching.

\paragraph{Shot Embedding.}
For each detected shot, we compute a 1024-dimensional visual embedding using Swin3D~\citep{liu2022swin3d}. Instead of processing the entire frames in a shot, we sample five representative frames at the 10th, 25th, 50th, 75th, and 90th percentiles and feed them together as a single clip into Swin3D to extract the shot-level embedding. This approach not only captures spatial information but also temporal changes within the shot, such as movements or camera motion, which helps differentiate one shot from the others in an ad.

\paragraph{Shot Clustering.}
Within each brand, we cluster shot embeddings using DBSCAN~\citep{ester1996density} with cosine distance, setting \textit{min\_samples} to 1 to allow singleton clusters for intentionally unique shots. We select $\varepsilon$ from 30 evenly spaced values in $[0.05, 0.9]$ via the first local maximum of the silhouette curve, yielding coherent clusters of visually similar shots while preserving unique ones as singletons.

\textbf{30sec-15sec Ad Pairs.} To determine which shots from the 30-sec videos appear in their 15-sec counterparts, we leverage the clustering results. Shots from the 30-sec and 15-sec ads that fall into the same cluster are considered semantically matched, reflecting the selections made by professional editors. 
These matched shots define the ground-truth shot selection sets, $S^*$, used for benchmarking MLLM performance. We retain a pair only when \emph{every} 15-sec shot matches a 30-sec shot ($S^* \subseteq \{s_1,\ldots,s_n\}$), so by construction no ground-truth shot is unmatched. This is what defines a cutdown rather than a simplification: a 15-sec shot absent from the master would require footage never shot for this ad, making it new creative rather than a cutdown, and would leave ground-truth shots no model could select. Ads with two or more released 15-sec versions are rare here and select very similar shots, so the released cutdown is a stable reference (\S\ref{sec:limitations}).%

\textbf{Dataset Accuracy.} To evaluate shot-matching quality, we randomly sampled 50 ad pairs for manual review. 
Two authors reviewed every pair independently and agreed on all 401 shot-level judgments (100\% inter-annotator agreement).%
 Across the 401 shots from the 50 15-sec ads, 394 were correctly identified and matched, resulting in an agreement of 98.2\% between the automated pair construction method and human judgment. 
The same review found 45 of the 50 15-sec ads preserve their source's shot order exactly.%


\textbf{Dataset Statistics.} AdShot contains 823 ad pairs (1,646 videos) from 194 unique brands spanning 17 industries. Table~\ref{tab:dataset_stats} presents industry-level statistics. The 30-sec source ads contain an average of 14 shots, while the professionally edited 15-sec ads include 8 shots on average, indicating that editors typically select around half of the original shots. Average shot durations are similar across the two versions (see Figure \ref{fig:shotstats}), suggesting that editors preserve temporal pacing rather than simply truncating content. The dataset covers 17 industries, capturing a wide range of visual styles, narrative structures, and pacing strategies used in professional advertising\footnote{To respect the intellectual property of the brands and comply with YouTube's Terms of Service, AdShot redistributes no source video. We release YouTube video IDs, shot-level temporal annotations, and content feature labels, together with derived per-shot keyframes and audio for the 30-sec source ads so the benchmark stays reproducible as videos disappear, under a non-commercial license (Appendix~\ref{app:availability}). Users can retrieve the full videos independently via the YouTube Data API in accordance with YouTube's Terms of Service.}.

\subsection{Shot-Level Content Features}
Each shot in an ad serves a strategic role, such as introducing the product, capturing attention, conveying brand information, or eliciting emotion. To analyze whether MLLMs select shots that align with professional editorial choices, we annotate each shot with content features reflecting these roles. Specifically, we use AdNotator~\citep{overgoor2025mis}, built on Video-LLaMA 2~\citep{damonlpsg2024videollama2}, which integrates vision and language understanding and has been fine-tuned on about 8,000 video ads for advertising-specific tasks. For each shot, AdNotator evaluates 18 binary features, which we organize into four high-level categories inspired by \cite{liaukonyte2015television}: action-focused (7 features, e.g., \textit{Is there a call to find out more information about the product or the brand (e.g., shop online, visit the Web)?} ), information-focused (3 features, e.g., \textit{Does the ad mention at least one specific product or service (e.g., model, type, item)?}), emotion-focused (5 features, e.g., \textit{Does the ad give you a positive feeling about the brand? }), and imagery-focused (3 features, e.g., \textit{Is the ad visually pleasing?}). These categories capture key advertising intents. A full list of the corresponding questions is provided in Appendix \ref{app:annotations}.

\tabdatasetstats

\section{Evaluation Methodology}
\subsection{Model Input and Output Format}
Our evaluation protocol provides MLLMs with multimodal inputs while constraining outputs for automated assessment. Figure~\ref{fig:evaluation} illustrates the full pipeline.

\textbf{Input Components.} Each model receives three types of information:
(1) \textit{Task instruction}: A system prompt specifying the task (select shots totaling $\sim$15s from a 30-sec ad), input format, and output requirements (see Appendix~\ref{app:prompts}  for full prompt). (2) \textit{Shot metadata}: Shot order and duration (seconds). (3) \textit{Visual frames}: For each shot, we extract frames at temporal percentiles (e.g., 50\%). Frames are provided sequentially with explicit shot boundary markers.

\figevaluation

\textbf{Output Format.} Models return structured JSON objects specifying: (1) \texttt{selected\_shots}: the list of chosen shot IDs, (2) \texttt{total\_length\_sec}: cumulative duration for constraint verification, and (3) \texttt{explanations}: reasoning for each selection.
\subsection{Evaluated Models}
We evaluate 13 video-only MLLMs and 2 audio-visual MLLMs spanning eight model families (7B--72B). Table~\ref{tab:models} summarizes the models; we briefly note each family's relevance to shot selection below.

\begin{table}[h]
\centering
\vspace{-1.0em}
\vspace{0.2em}
\small
\setlength{\tabcolsep}{12pt}
\begin{tabular}{llcc}
\toprule
\textbf{Model} & \textbf{Family} & \textbf{Params} & \textbf{Modality} \\
\midrule
LLaVA-OneVision~\citep{li2024llava}                  & LLaVA    & 7B, 72B & V   \\
Qwen2.5-VL~\citep{Qwen2.5-VL}                        & Qwen     & 7B, 32B & V   \\
Qwen3-VL~\citep{Qwen3-VL}                            & Qwen     & 8B, 32B & V   \\
InternVL2.5~\citep{chen2024expanding}                & InternVL & 8B, 26B & V   \\
InternVL3.5~\citep{zhu2025internvl3}                 & InternVL & 8B, 38B & V   \\
Aria~\citep{aria2024}                                & MoE      & 25.3B     & V   \\
MiniCPM-V~\citep{yu2025minicpm}                      & MiniCPM  & 8B      & V   \\
Molmo2~\citep{clark2026molmo2}                       & Molmo    & 8B      & V   \\
\midrule
Qwen2.5-Omni~\citep{xu2025qwen25omnitechnicalreport} & Qwen     & 7B      & A+V \\
MiniCPM-o 2.6~\citep{yao2024minicpmv}                & MiniCPM  & 8B      & A+V \\
\bottomrule
\end{tabular}
\caption{Evaluated models, parameter scales, and input modalities. V = video-only; A+V = audio + video.}
\label{tab:models}
\vspace{-0.8em}
\end{table}

\noindent Models were selected to cover complementary capabilities relevant to shot selection: \textbf{LLaVA-OneVision} offers strong single- and multi-image understanding; \textbf{Qwen2.5/3-VL} provides dynamic resolution and hour-long video processing with second-level localization; \textbf{InternVL2.5/3.5} pairs are included to measure version-to-version and scale-to-scale gains in temporal reasoning; \textbf{Aria} represents a MoE architecture trained on large-scale video-text data; \textbf{MiniCPM-V} benchmarks performance under compute constraints typical of editing workflows; \textbf{Molmo2} brings point-driven grounding and fine-grained temporal understanding as a specialist video model; and finally, \textbf{Qwen2.5-Omni} and \textbf{MiniCPM-o 2.6} are included as audio-visual models to evaluate the contribution of audio cues to shot selection.

\subsection{Evaluation Metrics}
We evaluate models on three dimensions: shot selection accuracy, duration constraint adherence, and content alignment.

\paragraph{Shot Selection Accuracy.}
We treat shot selection as set prediction, comparing model selections $S$ against ground truth $S^*$:
\begin{equation}
\text{Precision} = \frac{|S \cap S^*|}{|S|}, \quad
\text{Recall} = \frac{|S \cap S^*|}{|S^*|}, \quad
\text{F1} = \frac{2 \cdot \text{Precision} \cdot \text{Recall}}{\text{Precision} + \text{Recall}}.
\end{equation}
Precision captures over-selection (model picks shots that professional editors excluded), recall captures missing coverage (model omits shots the editors kept), and F1 balances the two.

\paragraph{Duration Constraint Adherence.}
To assess adherence to the 15s target (acceptable range 14--16s), we define the signed duration error $e = T - 15.0$, where $T = \sum_{i \in S} d_i$ is the total selected duration. We report the overall mean duration error along with the conditional mean magnitudes for undershoot and overshoot:
\begin{equation}
\text{MDE} = \mathbb{E}[|e|], \quad
\text{MDE}_{\text{neg}} = \mathbb{E}[|e| \mid e < 0], \quad
\text{MDE}_{\text{pos}} = \mathbb{E}[|e| \mid e > 0].
\end{equation}
Lower MDE indicates tighter adherence; comparing $\text{MDE}_{\text{neg}}$ and $\text{MDE}_{\text{pos}}$ reveals whether errors come from selecting too few shots or too many.

\paragraph{Content Alignment.}
Beyond selecting correct shots, we assess whether models preserve the content composition of professional edits. Let $f_i^j$ denote the output of AdNotator, where questions $A_c$ are posed based on category $c \in \{\text{Action, Information, Emotion, Imagery}\}$. Since the output is binary, $f_i^j \in \{0,1\}$. The average feature density $D_c$ is calculated as:
\begin{equation}
D_c(S) = \frac{1}{|S| \, |A_c|} \sum_{i \in S} \sum_{j \in A_c} f_i^j,
\end{equation}
where $A_c$ denotes the binary features in category $c$ and $|S|$ is the number of selected shots. We apply min-max normalization on $D_c(S)$, then measure the feature density difference between model selections $S$ and ground truth $S^*$:
\begin{equation}
\Delta_c = D_c(S) - D_c(S^*).
\end{equation}
Positive (negative) $\Delta_c$ indicates over- (under-) selection of features in category $c$ relative to professional editors, revealing systematic content biases.


\section{Experimental Results}
\label{sec:results}

\subsection{Settings}

\noindent\textbf{Video input format.}
Given a 30-sec ad video, we represent each shot by sampling its center frame (the 50th percentile). This single-frame-per-shot setting (1~FPS) balances computational efficiency with sufficient visual information, given the fact that shots only last ${\sim}2$ seconds on average. For audio-visual models, we additionally extract the audio track of each shot and provide it alongside the sampled frame.

\noindent\textbf{Inference configuration.}
We use greedy decoding for deterministic and reproducible results (\texttt{do\_sample=False}). All models are evaluated using their official implementations from HuggingFace Transformers or GitHub. Inference is conducted on NVIDIA A100 and H100 GPUs.

\tabperformance

\subsection{Shot Selection Performance}
Table~\ref{tab:main_results} presents overall performance across evaluated models. F1 scores range from 0.443 to 0.676\footnote{The ground-truth selections are professional human edits; an expert reproducing the original selection would achieve F1 = 1.0 by definition.}, with Aria (25.3B)\footnote{Aria's total parameter count is 25.3B; it activates 3.9B parameters per visual token and 3.5B per text token via its Mixture-of-Experts routing~\citep{aria2024}.} achieving the highest F1 of 0.676 through exceptional recall of 0.881 at the cost of lower precision of 0.576. The Qwen family demonstrates consistently competitive performance, with Qwen3-VL 32B achieving the best precision of 0.636 and Qwen2.5-VL 7B achieving the highest recall of 0.649 among Qwen models. Notably, larger models do not universally outperform smaller variants; for example, Qwen3-VL 8B (F1 = 0.576) underperforms both Qwen3-VL 32B (F1 = 0.611) and Qwen2.5-VL 7B (F1 = 0.582), suggesting architecture and training choices matter more than scale alone for this task. The audio-visual models, Qwen2.5-Omni 7B (F1 = 0.668) and MiniCPM-o 2.6 8B (F1 = 0.669), nearly match Aria 25.3B at roughly a third of the parameters, with similarly high recall but better precision (0.615 and 0.581 vs.\ 0.576). 
Removing the audio channel from these same two models lowers F1 to 0.585 and 0.644, driven by recall ($-0.178$ and $-0.063$), and the gain holds within industries (Appendix~\ref{app:audio_ablation}).%
 For reference, a heuristic baseline (Appendix~\ref{app:random_baseline}) that samples a random subset size and selects shots uniformly at random 
achieves F1 = 0.503. Nine of the fifteen models surpass it and six fall below, so extracting task-relevant signal here is model-dependent rather than a property of MLLMs in general --- the distinction AdShot exists to expose.

\paragraph{Models below the baseline.} The six (i.e., InternVL3.5 8B (0.494), Molmo2 8B (0.483), MiniCPM 8B (0.475), InternVL2.5 26B (0.473), LLaVA-OneVision 72B (0.464), InternVL3.5 38B (0.443)) share one signature: precision at or above the baseline (0.597--0.614 vs.\ 0.569) but recall far below it (0.366--0.450 vs.\ 0.541). They are conservative selectors, and since editors keep roughly half the source shots (8 of 14), F1 penalizes low recall heavily; all six correspondingly undershoot the 15s target. The failure is not one of scale. They span 8B to 72B and include the largest model evaluated, but of zero-shot fit: ad videos are dense and fast-cut, so models whose pretraining under-represents this style have a weaker basis for fine-grained selection. Closing that gap through fine-tuning on AdShot's pairs is left to future work (\S\ref{sec:limitations}).%

Table~\ref{tab:shotbin_results} further breaks down model performance by three target shot count categories ($<6, 6-10, >10$). Across most models, F1 scores are lowest for ads with fewer than 6 shots, reflecting the challenge of making precise selections when very few shots must be chosen. Performance improves for ads with 6--10 shots, with many models achieving their highest F1 in this intermediate range, suggesting that a moderate number of shots allows better alignment with professional editorial choices.
\paragraph{Trade-off.} We observe that models with higher precision than recall typically have $MDE_{\text{neg}} > MDE_{\text{pos}}$ (i.e., they undershoot the 15s target), while models with higher recall tend to have $MDE_{\text{pos}} > MDE_{\text{neg}}$ (i.e., they overshoot), suggesting a trade-off between precision--recall balance and meeting the duration budget. 
F1 and MDE are correspondingly correlated across models ($r = 0.78$), but they are computed independently: F1 measures only shot-set overlap and is neither rewarded nor penalized by landing in the 14--16s window. The correlation reflects a shared driver: selecting more shots raises both overlap and overshoot. Thus, selection quality and duration control remain two separable axes, and a useful model must satisfy both.%

\paragraph{Content biases.} Beyond shot accuracy, we analyze whether selected shots preserve the content composition of professional edits. Across all evaluated MLLMs, models systematically over-select emotion- and information-focused shots while under-selecting action and imagery; full results are documented in Appendix~\ref{app:content_alignment}.

\subsection{Duration Constraint Adherence}
We evaluate multimodal models on constrained shot selection. 
Averaged over the 13 video-only models, the mean duration error (MDE) is 5.39s (SD = 2.85), spanning $2.49$s to $11.95$s (Table~\ref{tab:main_results}); the two audio-visual models raise it to 6.09s. Either way the error is large against the 14--16s tolerance,%
 indicating difficulty in temporal planning and constraint control. Aria-25.3B exhibits the largest overshoot, with 11.95s, consistent with its high-recall strategy of selecting many shots. In contrast, some models show a tendency to undershoot, most notably InternVL3.5-38B ($MDE_{\text{neg}} =4.18$s). Overall, overshooting is slightly more common than undershooting; practically, overshoots are easier to trim, while undershoots can omit key content. The audio-visual models follow the same overshoot pattern as Aria (MDE = 11.18s and 10.06s for Qwen2.5-Omni 7B and MiniCPM-o 2.6 8B), reflecting their similarly high-recall selection behavior.

\tabshotbins


\section{Limitations and Future Work}
\label{sec:limitations}

\textbf{Task scope.} AdShot scores shot selection only. Reordering, intra-shot trimming, transitions, grading, and audio mixing are out of scope (\S\ref{sec:task}), and two are worth naming as unmeasured rather than absent. Shot \emph{order} and flow are not scored, since F1 is a set metric, though order is rarely what a cutdown changes (45 of 50 pairs preserve it); scoring it needs a different metric. Audio \emph{continuity} is not scored either: short-ad audio is often independently mixed, and reused audio still needs crossfading and re-timing.

\textbf{Duration coverage.} We cover the 30-to-15-sec adaptation, the most common pair of durations in television advertising~\citep{mediaradar2023}; shorter conversions (15$\to$10s, 10$\to$6s) and micro-ads remain unexplored, and our pipeline extends to them directly.

\textbf{Ground truth and annotations.} Each pair carries a single ground-truth edit, so an equally valid alternative selection is penalized like an incorrect one. We use the released cutdown because it is the edit a brand approved and aired, and multi-version cases are rare and highly similar; inter-editor variance is nonetheless unmeasured and would require commissioning independent professional edits. Content-bias findings depend on AdNotator~\citep{overgoor2025mis}, whose labels match the human-majority label on 79.88\% of a held-out 494-ad set, which is above human--human agreement (76.43\%), and agree with an independent frontier model on 81.96\% of 18,000 decisions on AdShot itself (Appendix~\ref{app:adnotator}). Agreement is lowest on the subjective Emotion and Imagery features, so bias estimates there carry the most uncertainty.

\textbf{Industry imbalance.} Legal (125 pairs) and Government (108 pairs) are over 28\% of the benchmark, while Electronics (2) and Apparel (4) are severely underrepresented---a reflection of the real ad market rather than curation. The per-industry breakdown (Appendix~\ref{app:industry}) shows F1 is stable across all 17 industries and the highest-volume ones sit near the overall average, so the imbalance does not drive the headline results; the smallest industries should not be read as per-domain findings.

\textbf{Future work.} The paired structure supports domain-specific fine-tuning, the most direct route to closing the gap for the sub-baseline models. Our ablations open two further axes: content-adaptive frame selection (choosing the most informative frame per shot rather than a fixed percentile) and content-aware audio processing, since raw unfiltered audio already yields measurable gains.%

\section{Conclusion}
We introduce AdShot, the first benchmark specifically designed to evaluate multimodal large language models on advertisement shot selection. Video advertisements are a uniquely challenging domain, requiring models not only to understand visual and narrative content but also to make editorial decisions that balance duration, composition, and persuasive intent. Our analysis demonstrates that, despite strong general video understanding, current MLLMs fall short of professional human standards, exhibiting systematic biases and difficulty in adhering to temporal constraints. By providing a rigorous evaluation framework grounded in real-world, professionally edited ads, AdShot enables precise assessment of both shot selection accuracy and content alignment, offering diagnostic insights that inform domain-specific fine-tuning and hybrid approaches integrating multimodal reasoning with advertising conventions. More broadly, this work highlights the importance of evaluating AI for strategic editorial decision-making, not only for comprehension.
\begin{ack}
\textbf{Funding.} This work received no direct third-party financial support.
Data collection was supported by API access credits from the YouTube Data API
credits program, which had no role in the design, analysis, or reporting of this
work.

\textbf{Competing interests.} The authors declare no competing interests.

\textbf{Data and code.} The AdShot benchmark (shot boundaries, matched 30-sec/15-sec pairs, and shot-level content-feature labels) and all evaluation code are available at \adshotrepo\ and \adshotdata; hosting, licensing, and Croissant metadata are documented in Appendix~\ref{app:availability}.
\end{ack}%

\clearpage
\bibliographystyle{plainnat}
\bibliography{ref}



\clearpage
\appendix

\section{Annotation Questions}
\label{app:annotations}
\begin{table}[!h]
\caption{Annotation Questions of Ad Content Feature by Category \label{tab:questions}}
\begin{tabular}{p{2cm} p{10cm}}
\toprule
\textbf{Category} & \textbf{Question} \\
\midrule
\textbf{Action} &
Q1: Is there a call to find out more information about the product or the brand (e.g., shop online, visit the Web)? \\
& Q2: Is there online contact information provided (e.g., URL, website)? \\
& Q3: Is there a visual or verbal call to purchase (e.g., buy now, order now)? \\
& Q4: Does the ad portray a sense of urgency to act (e.g., buy before sales ends, order before ends)? \\
& Q5: Is there an incentive to buy (e.g., a discount, a coupon, a sale or "limited time offer")? \\
& Q6: Is there offline contact information provided (e.g., phone, mail, store location)? \\
& Q7: Is there mention of something free? \\
\midrule
\textbf{Information} &
Q8: Does the ad mention at least one specific product or service (e.g., model, type, item)? \\
& Q9: Is there any verbal or visual mention of the price? \\
& Q10: Does the ad show the brand (logo, brand name) or trademark more than once? \\
\midrule
\textbf{Emotion} &
Q11: Is the ad intended to affect the viewer emotionally, with positive or negative emotion? \\
& Q12: Does the ad give you a positive feeling about the brand? \\
& Q13: Does the ad have a story arc, with a beginning and an end? \\
& Q14: Is the ad creative/clever? \\
& Q15: Is the ad intended to be funny? \\
\midrule
\textbf{Imagery} &
Q16: Does this ad provide sensory stimulation (e.g., cool visuals, arousing music)? \\
& Q17: Is the ad visually pleasing? \\
& Q18: Does the ad have cute elements like animals, babies, animated characters, etc? \\
\bottomrule
\end{tabular}
\end{table}

\clearpage

\section{System Prompts}
\label{app:prompts}

The complete system prompt provided to all models:

\begin{mdframed}[backgroundcolor=gray!10, linecolor=black, linewidth=0.5pt]
\small\ttfamily
TASK: Advertisement Shot Selection

You are an expert in advertisement clipping. Below is a 30-second video advertisement split into shots. Your task is to select a subset of whole shots whose total duration is approximately 15 seconds.

INPUT FORMAT:\\
- Each shot is represented by \{fps\} frames sampled at \{percentiles\} of each shot's duration\\
- Shots are labeled [Shot 1], [Shot 2], etc. with frames shown in order

OUTPUT FORMAT:\\
Return a JSON object with exactly these three fields:\\
\{\\
\hspace*{1em}'selected\_shots': ['shotX', 'shotY', 'shotZ',...], // Array of shot IDs\\
\hspace*{1em}'total\_length\_sec': XX.XX, // Sum of selected durations\\
\hspace*{1em}'explanations': ['reason for shotX', 'reason for shotY',...] // One per shot\\
\}

Example 1:\\
\{\\
\hspace*{1em}'selected\_shots': ['shot4', 'shot7', 'shot8', 'shot15'],\\
\hspace*{1em}'total\_length\_sec': 14.76,\\
\hspace*{1em}'explanations': ['reason for shot4', 'reason for shot7',\\
\hspace*{3em}'reason for shot8', 'reason for shot15']\\
\}

Example 2:\\
\{\\
\hspace*{1em}'selected\_shots': ['shot2', 'shot6', 'shot10'],\\
\hspace*{1em}'total\_length\_sec': 15.23,\\
\hspace*{1em}'explanations': ['reason for shot2', 'reason for shot6',\\
\hspace*{3em}'reason for shot10']\\
\}

TASK STEPS:\\
1. Review all shots and their durations provided below\\
2. Analyze the visual content of each shot\\
3. Select shots that together create an effective \textasciitilde15 second clip\\
4. Calculate total: sum the durations of all shots you selected\\
5. Verify your total is between 14-16 seconds\\
6. Write brief explanation for why each shot was selected

REQUIREMENTS:\\
- Select whole shots only (no partial shots)\\
- Total duration MUST be between 14-16 seconds (target: 15 seconds)\\
- total\_length\_sec MUST equal your actual calculated sum\\
- Return ONLY valid JSON with double quotes, no extra text
\end{mdframed}

\clearpage
\subsection*{User Prompt Format}
The user prompt combines shot metadata with the visual frames:

\begin{mdframed}[backgroundcolor=blue!5, linecolor=black, linewidth=0.5pt]
\small\ttfamily
This advertisement contains \{n\_shots\} shots. The duration of each shot (in seconds) is: \{"shot1": 1.25, "shot2": 2.10, "shot3": 1.84, ...\}

The frames of each shot are below:

{[Shot 1]}\\
{[Image 1]}\\
{[Image 2]}\\
{[Image 3]}\\

{[Shot 2]}\\
{[Image 4]}\\
{[Image 5]}\\
{[Image 6]}\\

...

{[Shot N]}\\
{[Image frames]}
\end{mdframed}

\noindent\textbf{Note:} The durations above are for illustration purposes. Each shot is preceded by a text label \texttt{[Shot i]} followed by its corresponding representative frames as images. The number of frames per shot depends on the FPS setting (1, 3, or 5 frames sampled at temporal percentiles).

\clearpage
\section{Content Alignment Analysis}
\label{app:content_alignment}
We randomly selected 415 pairs for content analysis to estimate the selection bias. Figure~\ref{fig:contentanalysis} reveals systematic preference across all evaluated video-only models. Overall, models over-select emotion-focused and information-rich shots while under-selecting action and imagery-driven content.

The divergence is most pronounced for information-focused shots, where models average 0.08 more informational features per shot than professional edits, suggesting prioritization of content pertaining to product/service, price, brand name/logo (see Appendix~\ref{app:annotations}) over others. Notably, model size does not mitigate these biases: Qwen3-VL-32B-Instruct and LLaVA-Onevision-72B exhibit similar patterns to 8B models, indicating that their decent shot selection performance (i.e., high F1 scores) can coexist with systematic content misalignment.

\figcontent

\subsection*{Ablation Study on Temporal Sampling}

In Figure~\ref{fig:allcomparision} (Corresponding table~\ref{tab:fps_performance}), we study denser temporal sampling at 1~FPS, 3~FPS using percentiles ${0.10, 0.50, 0.90}$ and at 5~FPS using percentiles ${0.10, 0.25, 0.50, 0.75, 0.90}$ to assess the impact of additional temporal context. Note that some models (e.g., InternVL2.5 and InternVL3.5) have input token limitations that prevent evaluation at 5 FPS.
Increasing sampling density (fps=3/5) tends to improve F1 for several models (e.g., Molmo2 and Qwen2.5-VL 7B), but it usually increases MDE, suggesting that additional temporal context helps models make better content-selection decisions while making it harder to satisfy strict duration constraints. Some models show no benefit or negative effects from denser sampling---most notably MiniCPM 8B. LLaVA-OneVision 7B is a mild positive exception, where MDE improves while F1 marginally increases.

\figcomparison
\tabfpsablation

\clearpage
\section{Heuristic Baseline}
\label{app:random_baseline}
To contextualize the absolute F1 and MDE scores in Table~\ref{tab:main_results}, we establish a chance-level reference that uses no visual, semantic, or duration information. Any model that meaningfully outperforms this baseline is exploiting structure beyond uniform random selection.

\subsection*{Formulation}
Given a 30-sec ad video with $n$ temporally ordered shots $\{s_1, \ldots, s_n\}$, the heuristic baseline operates in two steps:
\begin{enumerate}
    \item Sample a target selection size $N \sim \text{Uniform}\{1, 2, \ldots, n\}$.
    \item Draw $N$ distinct shot indices uniformly at random without replacement to form the selection set $S$.
\end{enumerate}
This baseline uses no information about the 15s duration target, shot content, or visual features. The variance in $N$ pushes selections toward both extremes (very few or very many shots) more frequently than the optimal $|S^*|$, which produces high MDE in both directions.

\subsection*{Setup and Results}
We evaluate across all 823 ad pairs in AdShot, averaging over 20 random seeds (standard deviations: $<$0.012 for F1/Precision/Recall, $<$0.3s for MDE). Reported values in Table~\ref{tab:main_results} are F1 = 0.503, Precision = 0.569, Recall = 0.541, MDE = 8.18s, MDE$_{\text{pos}}$ = 9.01s, MDE$_{\text{neg}}$ = 7.16s. The roughly balanced overshoot rate (55.3\%) and undershoot rate (44.5\%) confirm that the baseline is informationless: without any awareness of the duration budget, errors are distributed symmetrically around the 15s target.

The baseline's role is calibration, not competition: it fixes the F1 that a
selector using no visual, semantic, or duration information attains on this
benchmark, so the absolute scores in Table~\ref{tab:main_results} can be read
against a chance-level floor. Nine of the fifteen evaluated models exceed it and
six fall below it; \S\ref{sec:results} analyzes the shared low-recall signature
of the six. Note also that the baseline's recall (0.541) is comparatively high
precisely because sampling a uniform subset size often selects a large fraction
of the shots, recovering many ground-truth shots at the cost of a very large
duration error (MDE 8.18s) --- so a model that merely matches the baseline's F1
does not match its practical usefulness.

\newpage
\section{Per-Industry Performance}
\label{app:industry}

We use \textbf{Qwen3-VL~32B} as the representative model for this breakdown. In
Table~\ref{tab:main_results} it attains the best precision of any evaluated
model (0.636) together with the best balance of selection accuracy and duration
fidelity: a strong F1 (0.611) at a low duration error (MDE 3.71). Models with
higher recall or F1 reach those scores through heavy over-selection --- Aria
25.3B, for instance, attains 0.881 recall but an MDE of 11.95, more than three
times higher. Among models that keep the right \emph{amount} of content,
Qwen3-VL~32B is the most reliable, which makes it the natural single model to
characterize per-industry behavior.

Table~\ref{tab:industry} reports its Precision, Recall, and F1 for each of the
17 industries in the benchmark, sorted by F1. Shaded cells are above this
model's overall score for that metric; the overall row reproduces its row in
Table~\ref{tab:main_results}.

\begin{table}[htbp]
\centering
\footnotesize
\setlength{\tabcolsep}{9pt}
\caption{Per-industry shot-selection performance of Qwen3-VL~32B (1~FPS), sorted
by F1. Shaded cells are above the overall average (Precision~0.636,
Recall~0.618, F1~0.611). Industry labels follow Table~\ref{tab:dataset_stats}.}
\label{tab:industry}
\begin{tabular}{l r r r}
\toprule
\textbf{Industry} & \textbf{Prec.} & \textbf{Rec.} & \textbf{F1} \\
\midrule
Food \& Beverages   & \hi{0.707} & \hi{0.674} & \hi{0.674} \\
Cosmetic            & 0.618      & \hi{0.699} & \hi{0.634} \\
Restaurants         & \hi{0.664} & \hi{0.631} & \hi{0.634} \\
Services            & 0.619      & \hi{0.657} & \hi{0.625} \\
Health              & \hi{0.638} & \hi{0.641} & \hi{0.622} \\
Telecommunications  & 0.615      & \hi{0.661} & \hi{0.621} \\
Retail              & \hi{0.668} & 0.601      & \hi{0.616} \\
Travel              & \hi{0.647} & 0.602      & 0.611      \\
Automotive          & \hi{0.653} & 0.586      & 0.609      \\
Legal               & 0.624      & 0.608      & 0.601      \\
Home and Garden     & 0.627      & 0.589      & 0.598      \\
Consumer Products   & 0.611      & 0.591      & 0.588      \\
Government          & 0.610      & 0.595      & 0.584      \\
Education           & 0.608      & 0.592      & 0.572      \\
Entertainment       & 0.507      & \hi{0.662} & 0.551      \\
Apparel             & 0.609      & 0.447      & 0.513      \\
Electronics         & 0.486      & 0.517      & 0.498      \\
\midrule
\textbf{Overall}    & \textbf{0.636} & \textbf{0.618} & \textbf{0.611} \\
\bottomrule
\end{tabular}
\end{table}

Performance is stable across industries. The two lowest scores belong to the two
smallest industries in the benchmark (Electronics, $n{=}2$, and Apparel,
$n{=}4$; see Table~\ref{tab:dataset_stats}), whose 2--4-pair estimates are too
small to interpret. Every other industry falls within a narrow band (F1
0.551--0.674), with no domain collapsing.

This speaks directly to the industry imbalance discussed in
\S\ref{sec:limitations}. The three highest-volume industries --- Retail
($n{=}159$), Legal ($n{=}125$), and Government ($n{=}108$) --- all fall close to
the overall average (F1 0.616, 0.601, 0.584), so the dominant industries are not
outliers pulling the aggregate up or down, and the paper's headline numbers are
not an artifact of the industry mix. That no industry collapses also indicates
the shot-selection task is well defined across domains despite large differences
in volume, visual style, and pacing.

\newpage
\section{Audio Ablation}
\label{app:audio_ablation}

To isolate the contribution of audio, we re-evaluate the two audio-visual models
(Qwen2.5-Omni~7B and MiniCPM-o~2.6~8B) with the audio channel removed
(video-only frames, zero-shot, 1~FPS) and compare against their audio+video
results in Table~\ref{tab:main_results}. Because this is a same-model
comparison, any change is attributable to audio rather than to architecture or
scale.

\begin{table}[h]
\centering
\footnotesize
\setlength{\tabcolsep}{5pt}
\caption{Audio ablation: same model, same 823 AdShot pairs, with and without the
audio channel. Audio+Video rows are from Table~\ref{tab:main_results};
Video-only rows are this ablation. $\Delta$ is the audio contribution
(Audio+Video $-$ Video-only). MDE$_{\text{neg}}$ / MDE$_{\text{pos}}$ are the
mean under-/over-shoot in seconds.}
\label{tab:audio}
\begin{tabular}{l l r r r r r r}
\toprule
\textbf{Model} & \textbf{Input} & \textbf{Prec.} & \textbf{Rec.} & \textbf{F1} & \textbf{MDE} & \textbf{MDE$_{\text{neg}}$} & \textbf{MDE$_{\text{pos}}$} \\
\midrule
\multirow{3}{*}{MiniCPM-o 2.6 8B}
  & Audio+Video      & 0.581 & 0.841 & 0.669 & 10.06 & 1.72 & 10.50 \\
  & Video-only       & 0.588 & 0.778 & 0.644 & 8.38  & 2.28 & 9.60 \\
  & $\Delta$ (audio) & $-0.007$ & $+0.063$ & $\mathbf{+0.025}$ & $+1.68$ & $-0.56$ & $+0.90$ \\
\midrule
\multirow{3}{*}{Qwen2.5-Omni 7B}
  & Audio+Video      & 0.615 & 0.838 & 0.668 & 11.18 & 3.16 & 12.68 \\
  & Video-only       & 0.595 & 0.660 & 0.585 & 7.75  & 3.47 & 9.81 \\
  & $\Delta$ (audio) & $+0.020$ & $+0.178$ & $\mathbf{+0.083}$ & $+3.43$ & $-0.31$ & $+2.87$ \\
\bottomrule
\end{tabular}
\end{table}

Removing audio lowers F1 for both models --- by 0.025 for MiniCPM-o
(0.669$\to$0.644) and by 0.083 for Qwen2.5-Omni (0.668$\to$0.585) --- driven
mainly by a drop in recall ($-0.063$ and $-0.178$, respectively). Because these
are controlled same-model comparisons, they isolate audio's effect and upgrade
the paper's audio observation from a cross-architecture comparison to a direct
ablation: audio provides complementary signal that helps a model recover more of
the shots professional editors kept. The effect is consistent with the
precision/recall/duration trade-off in \S\ref{sec:results}: audio raises recall,
and in doing so lengthens the selection, raising the overshoot component of
duration error (MDE$_{\text{pos}}$ is higher with audio by 0.90s and 2.87s);
removing audio reduces MDE overall.

\paragraph{Per-industry audio contribution.}
To check that audio's benefit is not confined to a few domains,
Table~\ref{tab:audio_industry} repeats the Qwen2.5-Omni ablation within all 17
industries, sorted by volume. Audio raises recall in 16 of the 17 industries and
F1 in 14. The three F1 exceptions are minor: in Food~\& Beverages ($-0.019$) and
Restaurants ($-0.012$) audio's recall gain is offset by a small precision drop,
for a net change under 0.02; Apparel rests on only 4 ad pairs and is too small
to interpret, the same caveat as for the smallest industries in
Table~\ref{tab:industry}. The gain is consistently recall-driven, matching the
overall result.

\begin{table}[h]
\centering
\footnotesize
\setlength{\tabcolsep}{8pt}
\caption{Per-industry audio ablation for Qwen2.5-Omni~7B across all 17
industries, sorted by volume. ``$+$A'' is audio+video, ``$-$A'' is video-only;
$\Delta$ is the audio contribution. Shaded $\Delta$F1 cells are positive: audio
improves F1 in 14 of 17 industries and recall in 16 of 17.}
\label{tab:audio_industry}
\begin{tabular}{l r r r r r r}
\toprule
\textbf{Industry} & \textbf{F1$_{+\text{A}}$} & \textbf{F1$_{-\text{A}}$} & \textbf{$\Delta$F1}
 & \textbf{Rec$_{+\text{A}}$} & \textbf{Rec$_{-\text{A}}$} & \textbf{$\Delta$Rec} \\
\midrule
Retail              & 0.659 & 0.586 & \hi{$+0.073$} & 0.825 & 0.639 & $+0.186$ \\
Legal               & 0.644 & 0.602 & \hi{$+0.042$} & 0.822 & 0.689 & $+0.133$ \\
Services            & 0.648 & 0.567 & \hi{$+0.082$} & 0.883 & 0.716 & $+0.167$ \\
Government          & 0.669 & 0.563 & \hi{$+0.106$} & 0.825 & 0.611 & $+0.213$ \\
Health              & 0.640 & 0.575 & \hi{$+0.064$} & 0.817 & 0.637 & $+0.180$ \\
Automotive          & 0.727 & 0.571 & \hi{$+0.156$} & 0.920 & 0.604 & $+0.316$ \\
Food \& Beverages   & 0.621 & 0.640 & $-0.019$      & 0.742 & 0.721 & $+0.022$ \\
Telecommunications  & 0.664 & 0.627 & \hi{$+0.037$} & 0.820 & 0.689 & $+0.131$ \\
Education           & 0.697 & 0.584 & \hi{$+0.113$} & 0.935 & 0.699 & $+0.236$ \\
Travel              & 0.668 & 0.580 & \hi{$+0.088$} & 0.806 & 0.650 & $+0.156$ \\
Consumer Products   & 0.660 & 0.560 & \hi{$+0.101$} & 0.854 & 0.620 & $+0.234$ \\
Restaurants         & 0.651 & 0.663 & $-0.012$      & 0.772 & 0.742 & $+0.030$ \\
Home and Garden     & 0.696 & 0.549 & \hi{$+0.148$} & 0.835 & 0.574 & $+0.261$ \\
Entertainment       & 0.602 & 0.552 & \hi{$+0.049$} & 0.821 & 0.741 & $+0.080$ \\
Cosmetic            & 0.610 & 0.602 & \hi{$+0.008$} & 0.796 & 0.717 & $+0.079$ \\
Apparel             & 0.287 & 0.444 & $-0.157$      & 0.224 & 0.527 & $-0.303$ \\
Electronics         & 0.671 & 0.557 & \hi{$+0.114$} & 0.642 & 0.508 & $+0.133$ \\
\bottomrule
\end{tabular}
\end{table}

\newpage
\section{Qualitative Examples}
\label{app:qual}

\begin{figure}[h]
\centering
\includegraphics[width=0.8\linewidth]{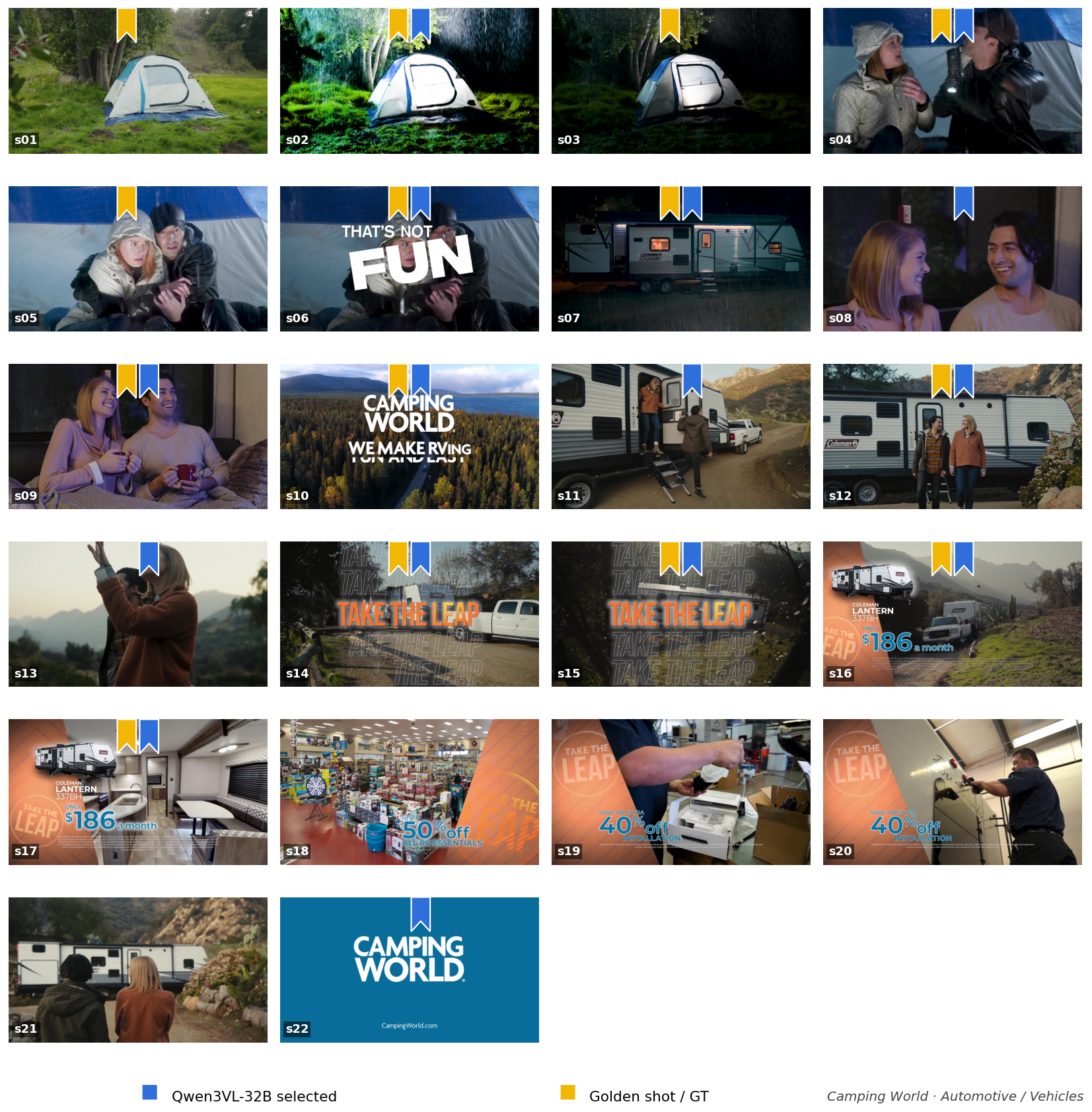}
\caption{Example 1 --- Camping World (Automotive / Vehicles), 22 shots. Blue:
Qwen3-VL~32B selected; yellow: golden shot (ground truth).}
\label{fig:qual1}
\end{figure}

\begin{figure}[h]
\centering
\includegraphics[width=0.8\linewidth]{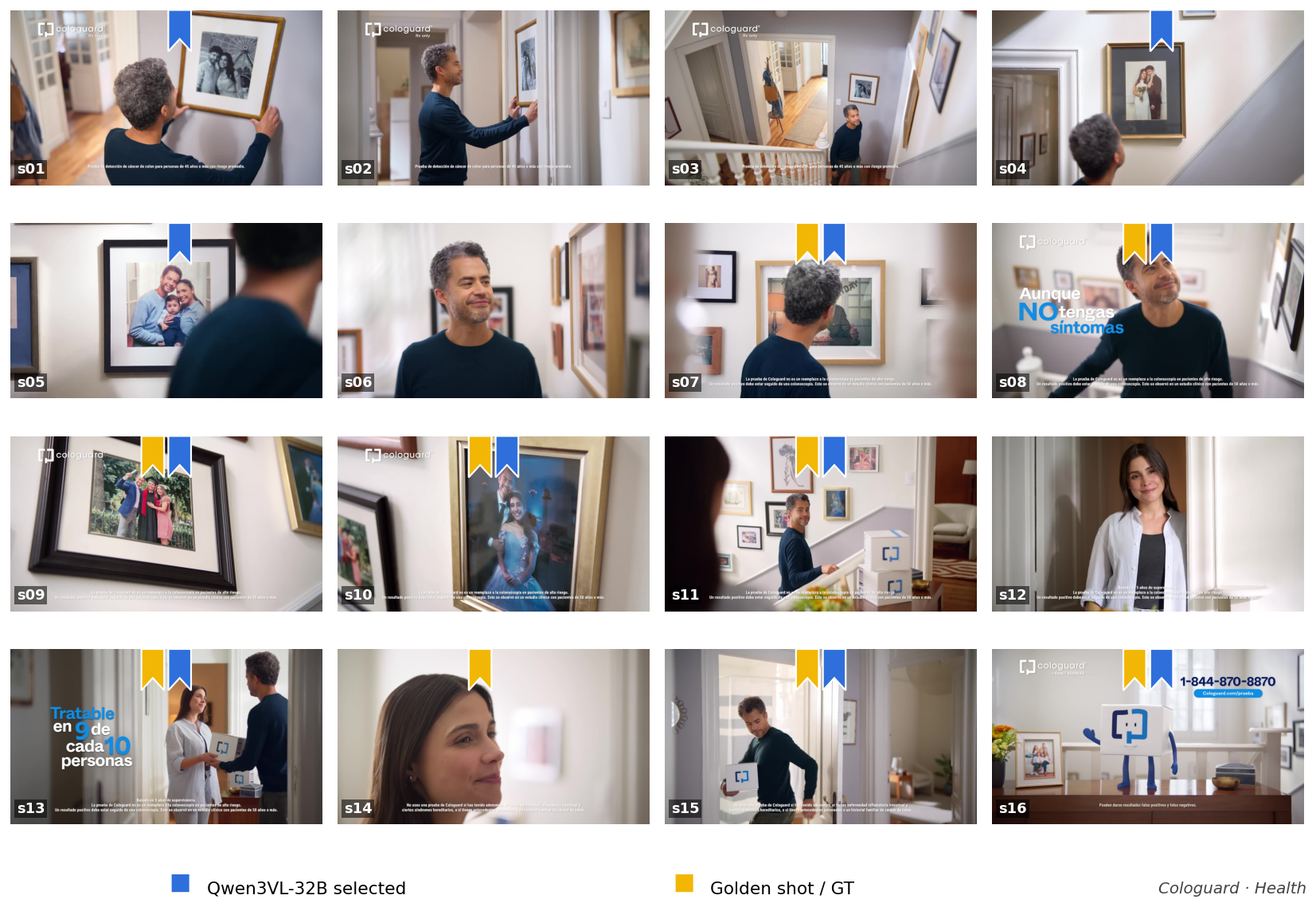}
\caption{Example 2 --- Cologuard (Health), 16 shots. Blue: Qwen3-VL~32B
selected; yellow: golden shot (ground truth).}
\label{fig:qual2}
\end{figure}

\begin{figure}[h]
\centering
\includegraphics[width=0.8\linewidth]{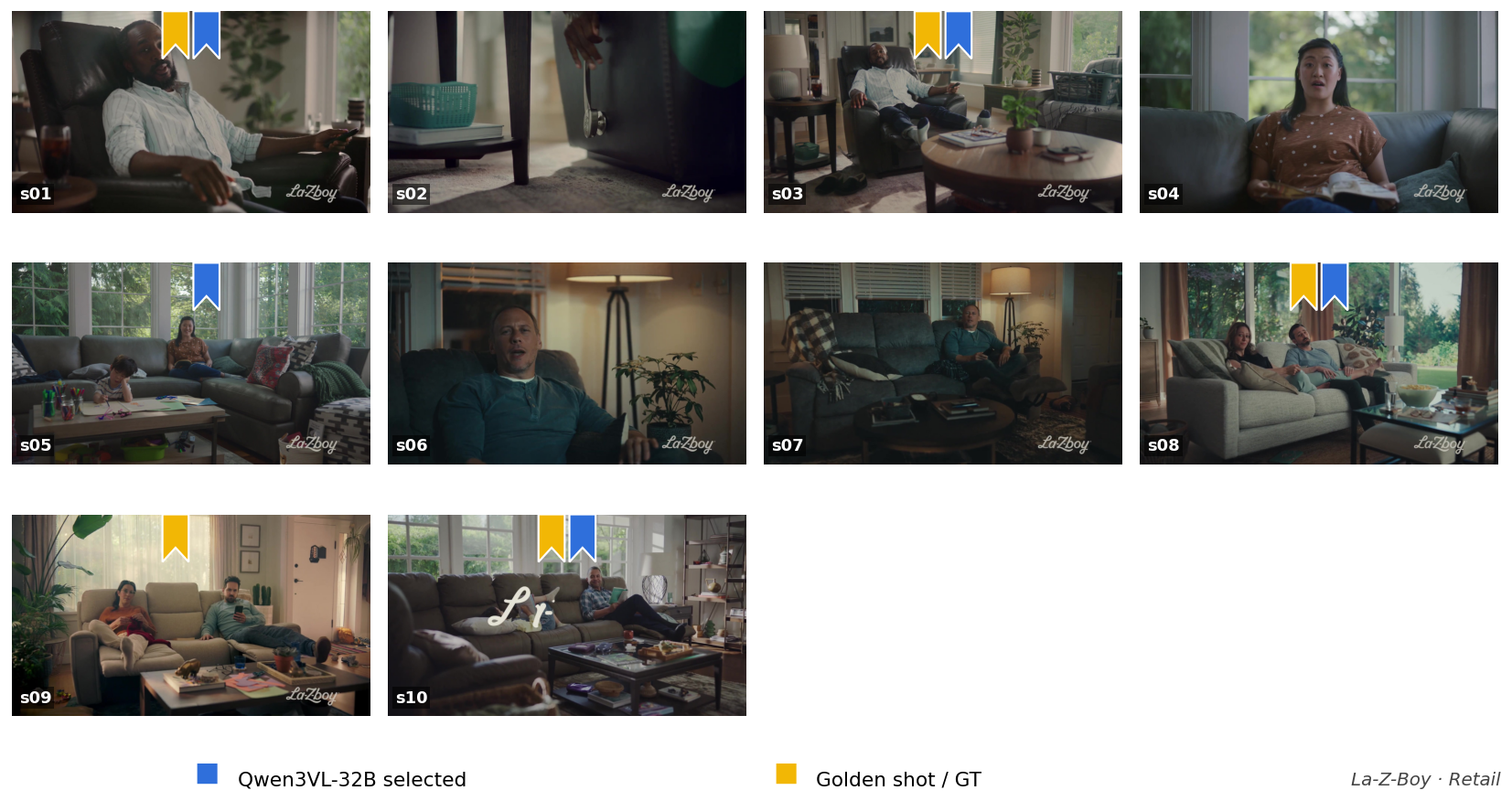}
\caption{Example 3 --- La-Z-Boy (Retail), 10 shots. Blue: Qwen3-VL~32B selected;
yellow: golden shot (ground truth).}
\label{fig:qual3}
\end{figure}

Figures~\ref{fig:qual1}--\ref{fig:qual3} show three complete ads, each shot
displayed in order and represented by a single frame for readability, so an
entire ad fits in one figure. The three span different brands, industries, and
lengths: Camping World (Automotive, 22 shots), Cologuard (Health, 16 shots), and
La-Z-Boy (Retail, 10 shots). A \textbf{blue} marker denotes a shot selected by
Qwen3-VL~32B; a \textbf{yellow} marker denotes a golden (ground-truth) shot kept
by the professional editor. A shot with both markers is a correct match;
blue-only is an over-selection (false positive); yellow-only is a miss (false
negative).

Across the three ads the model performs genuine selection rather than keeping
everything: it leaves out shots the editor also dropped (no marker), for example
the in-store and offer shots in Camping World (s18--s20). It recovers the shots
that anchor each edit --- the brand and offer frames the editor kept (both
markers; Camping World s10 and s14--s17, Cologuard s16, La-Z-Boy s10). Its
errors are systematic rather than random and skew toward over-selection rather
than omission: it adds a few shots the editor dropped (blue-only; e.g.\ Camping
World s08 and s13, the emotional couple shots) somewhat more often than it
misses a shot the editor kept (yellow-only; Camping World s01/s03, Cologuard
s14, La-Z-Boy s09 --- mostly establishing or lifestyle shots). This
over-selection skew is consistent with the positive duration error this model
shows in Table~\ref{tab:main_results} ($\text{MDE}_{\text{pos}}$ 3.31 vs.\
$\text{MDE}_{\text{neg}}$ 0.39). The three ads span different industries and
sizes (Automotive 22 shots, Health 16, Retail 10), yet the same pattern holds.

\newpage
\section{Frontier-Model Agreement with AdNotator}
\label{app:adnotator}

AdShot's content-bias analysis relies on AdNotator~\citep{overgoor2025mis} for
shot-level content labels. AdNotator is human-validated in that work, reaching
79.88\% agreement with the human-majority label on a held-out 494-ad set ---
above human--human agreement (76.43\%) and above Gemini~2.5 Flash (75.41\%). As
additional evidence that these labels are reasonable on AdShot specifically, we
checked whether a strong, independent frontier model reproduces them.

\paragraph{Process.} We drew a random sample of 1{,}000 AdShot shots and labeled
each with Gemini~3.1 Pro on the same 18 binary content-feature questions
AdNotator uses (Appendix~\ref{app:annotations}), grouped into the four
categories Action, Information, Emotion, and Imagery. Comparing the two label
sets question by question yields 18{,}000 matched question--shot decisions.

\paragraph{Results.} Gemini~3.1 Pro agrees with AdNotator on 81.96\% of the
18{,}000 decisions (Table~\ref{tab:adnotator}). Agreement is highest on the
Action features (93.96\%) --- which turn on concrete, verifiable cues such as a
displayed URL, a price, or a stated offer --- and lower on the more subjective
Emotion (76.44\%) and Imagery (70.63\%) features, as expected for judgments on
which even human annotators diverge. This is supplementary context rather than a
substitute for the human validation above: it shows that a strong independent
model largely reproduces AdNotator's labels, not which of the two is more
correct. AdShot's content-bias results continue to rely on AdNotator as the
trained, human-validated source, and \S\ref{sec:limitations} notes that
category-level bias estimates for Emotion and Imagery carry the most
uncertainty.

\begin{table}[h]
\centering
\footnotesize
\setlength{\tabcolsep}{10pt}
\caption{Agreement between Gemini~3.1 Pro and AdNotator on a random sample of
1{,}000 AdShot shots, by content-feature category and overall. Each shot is
labeled on the same 18 binary questions (7 Action, 3 Information, 5 Emotion, 3
Imagery), giving 18{,}000 matched question--shot decisions.}
\label{tab:adnotator}
\begin{tabular}{l r r}
\toprule
\textbf{Category} & \textbf{\# Questions} & \textbf{Agreement (\%)} \\
\midrule
Action       & 7  & 93.96 \\
Information  & 3  & 74.47 \\
Emotion      & 5  & 76.44 \\
Imagery      & 3  & 70.63 \\
\midrule
\textbf{Overall} & \textbf{18} & \textbf{81.96} \\
\bottomrule
\end{tabular}
\end{table}

\newpage
\section{Dataset Documentation, Licensing, and Availability}
\label{app:availability}


\paragraph{What we release.} We release no source video. For each of the 823 ad
pairs we distribute an annotation table containing the YouTube video IDs of the
30-sec source and its 15-sec cutdown, shot-level temporal boundaries for both
videos, the ground-truth matching between 15-sec and 30-sec shots ($S^*$), the
18 binary content-feature labels per shot, and brand and industry metadata. To
keep the benchmark runnable as source videos disappear from YouTube, we
additionally release two sets of \emph{derived} per-shot media for the 30-sec
source ads only: RGB keyframes (five per shot, sampled at the 10th, 25th, 50th,
75th, and 90th percentiles of each shot, matching the densest setting of our
temporal-sampling ablation) and 16~kHz mono audio waveforms. Keyframes are
available for 819 of the 823 source ads; four are missing because the source
videos were removed, made private, or age-gated before extraction. No media is
released for the 15-sec target ads, which serve only as the source of the
ground-truth labels. We also release the full evaluation harness --- prompt
templates, per-model inference scripts, and the metric implementations for
Precision/Recall/F1, MDE, and content-feature density --- together with the
per-model raw outputs needed to reproduce every table in this paper.

\paragraph{Hosting and persistence.} The dataset is hosted on Hugging Face at
\adshotdata, and the code at \adshotrepo. Releasing derived
per-shot frames and audio rather than whole videos follows the practice of
large-scale YouTube-sourced datasets, which distribute identifiers and derived
annotations rather than the source media (Kinetics, HowTo100M, AVA); users who
need the complete videos retrieve them themselves via the YouTube Data API. The
derived media is what makes the benchmark robust to link rot: because the model
inputs our evaluation actually consumes are the per-shot frames and audio, the
benchmark remains fully reproducible even after source videos become
unavailable, and the annotation layer --- boundaries, pair matchings, and
content-feature labels --- remains usable independently. We periodically
re-validate video ID availability and publish the attrition rate alongside the
release, and we maintain the dataset for corrections, issues, and takedown
requests.

\paragraph{Licensing.} The annotations, derived media, and metadata are released
under CC~BY-NC~4.0, which permits reuse and redistribution for non-commercial
research with attribution; the evaluation code is released under the MIT
license. The underlying advertisements remain the property of their respective
brands and are not licensed by us; users are responsible for any use of the
source videos in accordance with YouTube's Terms of Service. The non-commercial
term is deliberate: it keeps the release scoped to research use of third-party
commercial creative. AdShot contains no personally identifiable information.
The videos are commercial advertisements published on official brand channels,
and we release no viewer, user, or channel-subscriber data; people appearing in
the ads, and their voices in the released audio, are incidental to the
advertising content and are not the subject of any annotation.

\paragraph{Croissant metadata.} Following the NeurIPS Evaluations \& Datasets
requirements, the release includes a validated Croissant file (conforming to
Croissant~1.1) covering both the core fields --- dataset name, URL, license,
version, citation, the three distributed files, and the \texttt{RecordSet}
structure of the ad-pair records --- and the Responsible~AI fields: data
collection and collection type, annotation and preprocessing protocols, data
biases and limitations (\S\ref{sec:limitations}, including the industry
imbalance, the single-edit ground truth, and the US-market and English-language
skew), personal and sensitive information (none), intended use cases
(evaluating shot selection for duration cutdowns; not intended for training
production ad-generation systems without human oversight), social impact,
synthetic data (none --- all ads are professionally produced and publicly
aired), and the release and maintenance plan.
\end{document}